\documentclass[lettersize,journal]{IEEEtran}
\usepackage{amsmath,amsfonts}
\usepackage{algorithmic}
\usepackage{algorithm}
\usepackage{xcolor}
\usepackage{array}
\usepackage{subcaption}
\usepackage{textcomp}
\usepackage{stfloats}
\usepackage{url}
\usepackage{verbatim}
\usepackage{graphicx}
\def\BibTeX{{\rm B\kern-.05em{\sc i\kern-.025em b}\kern-.08emT\kern-.1667em\lower.7ex\hbox{E}\kern-.125emX}}
\usepackage{balance}
\usepackage{cite}
\usepackage{xurl}

\begin{document}

\title{Leveraging Compression to Construct Transferable Bitrate Ladders}

\author{
    Krishna Srikar Durbha, Hassene Tmar, Ping-Hao Wu, Ioannis Katsavounidis and Alan C. Bovik,~\IEEEmembership{Life Fellow,~IEEE}
    \thanks{
        This research was sponsored by a grant from Meta Media Foundation, and by grant number 2019844 from the National Science Foundation AI Institute for Foundations of Machine Learning (IFML).
    }
    \thanks{
        Krishna Srikar Durbha is with the Laboratory for Image and Video Engineering (LIVE), The University of Texas at Austin, Austin, TX 78712, USA (Email: krishna.durbha@utexas.edu)
    }
    \thanks{
        Alan C. Bovik is with the Colorado's Laboratory for Image and Video Engineering (LIVE), University of Colorado Boulder, Boulder, CO 80309 USA (e-mail: Alan.Bovik@colorado.edu).
    }
    \thanks{
        Hassene Tmar, Ping-Hao Wu, and Ioannis Katsavounidis are with Meta Platforms Inc., Menlo Park, CA 94025, USA (Email: htmar@meta.com; npw@meta.com; ikatsavounidis@meta.com).
    }
    \thanks{
        All databases accessed during training and evaluation were performed at The University of Texas at Austin by university faculty, staff, and students. The implementations used in this work have been made available at https://github.com/krishnasrikard/Leveraging-Compression-to-Construct-Transferable-Bitrate-Ladders.
    }
}

\markboth{Journal of \LaTeX\ Class Files, ~Vol.~xx, No.~x, Month~20xx}{}

\maketitle

\begin{abstract}
    Over the past few years, per-title and per-shot video encoding techniques have demonstrated significant gains as compared to conventional techniques such as constant CRF encoding and the fixed bitrate ladder. These techniques have demonstrated that constructing content-gnostic per-shot bitrate ladders can provide significant bitrate gains and improved Quality of Experience (QoE) for viewers under various network conditions. However, constructing a convex hull for every video incurs a significant computational overhead. Recently, machine learning-based bitrate ladder construction techniques have emerged as a substitute for convex hull construction. These methods operate by extracting features from source videos to train machine learning (ML) models to construct content-adaptive bitrate ladders. Here, we present a new ML-based bitrate ladder construction technique that accurately predicts the VMAF scores of compressed videos, by analyzing the compression procedure and by making perceptually relevant measurements on the source videos prior to compression. We evaluate the performance of our proposed framework against leading prior methods on a large corpus of videos. Since training ML models on every encoder setting is time-consuming, we also investigate how per-shot bitrate ladders perform under different encoding settings. We evaluate the performance of all models against the fixed bitrate ladder and the best possible convex hull constructed using exhaustive encoding with Bj{\o}ntegaard-delta metrics.
\end{abstract}


\section{Introduction}
\label{sec:introduction}
A recent report on video streaming \cite{Video-Streaming-Report} predicts that approximately 75\% of internet traffic is composed of video content and is projected to reach 80\% by 2029. Substantial increases in the volume of User-Generated Content (UGC) on social media platforms and demand for Video-on-Demand (VoD) services offered by Netflix, Meta, YouTube, and Prime Video are driving video service providers to develop delivery systems that can adapt to highly diverse and varying network conditions, user preferences, display settings, and other factors. Videos stored on streaming platforms undergo compression under perceptual constraints to maximize the Quality of Experience (QoE) for users. Video service providers have invested considerable resources to improve video compression and delivery pipelines, to reduce bandwidth consumption, video transmission costs, and to improve end-user satisfaction.

In recent years, HTTP Adaptive Streaming (HAS) has gained popularity as a standard for transmitting videos to users under varying conditions. In particular, the HTTP Live Streaming bitrate ladder \cite{Fixed-Bitrate-Ladder} has emerged as a conventional and straightforward solution for providing adaptive bitrate streaming across various types of video content. The fixed bitrate ladder, also referred to as “one-size-fits-all,” is intended to accommodate diverse video characteristics and network conditions. However, the static nature of the “one-size-fits-all” approach over all kinds of video content can lead to sub-optimal streaming performance and reduced Quality of Experience (QoE). Per-shot (or per-scene) encoding techniques and the Dynamic Optimizer \cite{Per-title-Encoding, Shot-Encoding, Dynamic-Optimizer, Iterative-techniques-for-encoding-video-content} offer enhancements of video quality and improved bitrate savings, as compared to conventional fixed bitrate ladder approaches. Per-shot encoding techniques optimize the QoE by employing distinct operating points for each shot or scene in a video. Each video to be delivered is divided into multiple shots or scenes of relatively short durations, that are encoded independently of each other. To determine optimal operating points, each shot is compressed at various resolutions and rate-control settings. A convex hull is constructed for each shot by estimating the perceptual quality of each compressed shot using a full-reference video quality assessment algorithm (FR-VQA). However, a notable challenge associated with this process are the substantial resources and time required to find optimal operating points over a large number of video shots. For example, consider a video streaming scenario allowing $\mathcal{R}$ spatial resolutions and $\mathcal{B}$ rate-control settings (CRFs, QPs, or bitrates). Each shot of a video must be compressed and quality estimated $\mathcal{R} \times \mathcal{B}$ times to construct the convex hull. The execution time (system + user) of compression depends on the encoder settings, such as resolution, CRF/QP, preset, etc. Additionally, the quality estimation process depends on the FR-VQA algorithm used, such as SSIM \cite{SSIM}, VIF \cite{VIF}, VMAF \cite{VMAF}, and so on. Throughout this paper, the terms `shot' and `video' will be used interchangeably.

Over the past few years, various authors have proposed efficient ways of constructing per-shot bitrate ladders using machine learning techniques. Early work focused on employing fast encoder implementations, while more recent efforts have employed analytical, experimental, or machine learning techniques to predict content-aware per-shot bitrate ladders with significant emphasis on time constraints and computational expense. Most of these methods have involved design frameworks whereby various spatial and temporal features are computed on exemplar source videos to construct per-shot bitrate ladders by training regressors to predict bitrate, quality, cross-over QPs, knee-points, and more. However, both categories of methods present certain disadvantages. While constructing a per-shot convex hull using fast encoder settings provides significant improvements relative to the original idea of independently constructing a convex hull for each encoder setting, this procedure still requires expending significant resources on convex hull construction for each shot. Conversely, while ML-based bitrate ladder construction methods can offer enormous improvements in construction speeds, a significant issue with these methods is that ML models need to be extensively trained on large samples of highly diverse videos before they can be deployed in real-world scenarios. Also, to the best of our knowledge, per-shot bitrate ladders constructed using ML-based methods have not been studied and tested across different encoders and their settings. The methods we describe offer distinct advantages in terms of performance and reduced complexity, as compared to prior ML-based techniques and compute-intensive processes involved in constructing convex hulls, respectively.

\subsection{Contributions}
In preliminary work \cite{Bitrate-Ladder-Construction-using-Visual-Information-Fidelity} presented in a conference paper, we experimented with features drawn from Visual Information Fidelity (VIF) to predict the quality of videos compressed at a given resolution and bitrate. We presented multiple VIF feature sets extracted over different scales and subbands of a video, and used them to train a regressor to predict the quality of compressed videos using VIF features from given source videos, and the target bitrates, widths, and heights of the compressed videos. Subsequently, a trained regressor is used to construct per-shot bitrate ladders. In our recent work \cite{Constructing-Per-Shot-Bitrate-Ladders-using-Visual-Information-Fidelity}, we greatly improved upon this preliminary work by using the VIF features to construct per-shot video quality ladders along with bitrate ladders. We also proposed an ensemble of content features: low-level video features and VIF quality features, which we used to construct per-shot content-adaptive bitrate and quality ladders. We designed multiple regression models trained on these content-adaptive features, along with metadata such as frame width and height, and bitrate or quality, to predict the quality or bitrate of compressed videos, respectively. These prediction models were then used to construct per-shot bitrate ladders with bitrate steps, or quality ladders with VMAF steps. We have since greatly expanded on these directions of work, making the following new contributions:
\begin{itemize}
    \item We propose a novel bitrate ladder construction framework that efficiently predicts the VMAF scores of compressed videos, by leveraging various compression statistics or logs obtained by compressing videos and using various features extracted from source videos.
    \item We study the performance of our proposed framework against prior bitrate ladder construction methods, on a comprehensive dataset of videos. Furthermore, we assess the transferability of bitrate ladders constructed for fast encoder settings across different codecs and presets.
    \item We compare the performance of the models constructed using our framework along with prior state-of-the-art methods, against the fixed bitrate ladder and the convex hull constructed using exhaustive encoding across various encoder settings and employing Bj{\o}ntegaard-delta metrics.
\end{itemize}

\subsection{Paper Organization}
The rest of the paper is organized in the following structure. Section \ref{sec:related_works} discusses prior work related to the construction of per-shot bitrate ladders. Section \ref{sec:method} introduces our novel bitrate ladder construction framework and various features employed in our experiments, including quality-aware VIF feature sets and content-aware low-level features. Section \ref{sec:dataset-experimental-settings} discusses the dataset and experimental settings. Section \ref{sec:experiments} describes the experiments we conducted and Section \ref{sec:results} presents the performance of all the compared models. The paper concludes in Section \ref{sec:conclusion}.

\section{Related Work}
\label{sec:related_works}
As mentioned in the Introduction, despite its user-friendly nature, the fixed bitrate ladder \cite{Fixed-Bitrate-Ladder}, even when designed considering diverse video characteristics and network conditions, results in subpar reports or measurements of Quality of Experience (QoE). The per-shot encoding framework \cite{Per-title-Encoding, Shot-Encoding, Dynamic-Optimizer,Iterative-techniques-for-encoding-video-content} introduced by Netflix promotes exhaustive encoding of source videos across multiple resolutions and diverse rate-control settings. A comprehensive search for the optimal parameter settings is conducted within the encoding parameter space, using state-of-the-art full reference video quality assessment algorithms (FR-VQA) as an evaluation metric. The convex hull, constructed as a result, describes the optimal operating rate-quality (RQ) points of a given video for a set of given encoding settings.

Instead of relying on the exhaustive process of convex hull construction for different sets of encoder settings (e.g., codec and preset pair), the authors of \cite{Fast-Encoding-Parameter-Selection-for-Convex-Hull-Video-Encoding} suggest constructing the convex hull on faster encoder settings. During the encoding process, the optimal encoding parameters obtained with a faster encoder setting are directly applied to a slower encoder setting. The authors demonstrated significant reductions in computation complexity. The same authors further extended their work \cite{Encoding-Parameters-Prediction-for-Convex-Hull-Video-Encoding} by developing a predictive model to compensate for the loss incurred by employing the fast encoder. The authors employed a regressor to predict bitrate and quality pairs of slow encoders, from the outcomes of faster encoder, across a fixed coding standard, and by mapping QPs when predicting across different coding standards. Their proposed techniques reduce the losses compared to their previous work \cite{Fast-Encoding-Parameter-Selection-for-Convex-Hull-Video-Encoding} with slight increases in computation.

More recent methods employ machine learning techniques to construct per-shot, content-dependent bitrate ladders. In \cite{Predicting-Video-Rate-Distortion-Curves-using-Textural-Features}, the authors model PSNR as a linear function of the logarithm of bitrate. A support vector regressor (SVR) is trained on features including gray-level co-occurrence matrices \cite{GLCM} (GLCM), temporal coherence (TC) \cite{TC}, and normalized cross-correlations (NCC) to predict the coefficients of a linear model. They extended their work in \cite{Study-of-compression-statistics-and-prediction-of-rate-distortion-curves-for-video-texture} using an expanded feature set and by deploying various parametric functions to fit Rate-Quality (RQ) curves, including linear, second-order, third-order, and exponential functions. They showed that while third-degree polynomials provided close fits to rate-quality (RQ) curves, using an exponential function yielded superior BD rate savings with reduced complexity. The authors of \cite{Content-gnostic-Bitrate-Ladder-Prediction-for-Adaptive-Video-Streaming} and \cite{Efficient-Bitrate-Ladder-Construction-for-Content-Optimized-Adaptive-Video-Streaming} introduce a content-driven bitrate ladder prediction method that predicts the cross-over points between Rate-PSNR curves. At each cross-over point, the optimal resolution is switched from a lower resolution to a higher resolution as the bitrate increases. They define cross-over points as pairs of QPs corresponding to each resolution. Low-level features like GLCM and TC are used to predict the cross-over QPs. In \cite{VMAF-based-Bitrate-Ladder-Estimation-for-Adaptive-Streaming}, the authors predicted the bitrate ladder by predicting the `knee-points' of RQ curves, defined as these QPs at which the rate-quality curve has the highest curvature. Low-level features like the GLCM and NCC are used to predict the knee QPs.

Unlike \cite{Content-gnostic-Bitrate-Ladder-Prediction-for-Adaptive-Video-Streaming} and \cite{Efficient-Bitrate-Ladder-Construction-for-Content-Optimized-Adaptive-Video-Streaming}, where the authors model cross-over points as QPs, the authors of \cite{Benchmarking-Learning-based-Bitrate-Ladder-Prediction-Methods-for-Adaptive-Video-Streaming} model them as bitrates. To predict cross-over bitrates between consecutive resolutions, they experimented with various shallow ML models including Extra-Trees regressor \cite{Extra-Trees}, XGBoost \cite{XGBoost}, and Gaussian Process \cite{GP} regression, trained on low-level features like GLCM, TC, spatial information (SI) \cite{SI-TI}, temporal information (TI) \cite{SI-TI}, and Colorfulness (CF) \cite{CF}, along with features drawn from semantic-aware deep learning models like VGG16 \cite{VGG} and ResNet50 \cite{ResNet}. They reported that the Extra-Trees regressor delivered the best performance. In \cite{Perceptually-Aware-Per-Title-Encoding-for-Adaptive-Video-Streaming}, the authors predicted content-adaptive bitrate ladders by employing a quality prediction model. They modeled VMAF as a linear function of bitrate and DCT-based texture energy features, obtaining good correlations between true and predicted VMAF scores. They extended their work in \cite{Just-Noticeable-Difference-aware-Per-Scene-Bitrate-laddering-for-Adaptive-Video-Streaming} to predict VMAF scores, optimal CRFs, and Just Noticeable Difference (JND) thresholds using the same DCT-based energy features, luminescence, and bitrate. The number of predicted encoding settings was further reduced during bitrate ladder construction using JND thresholds predicted using GLCM features, bitstream features, etc. In \cite{Ensemble-Learning-for-Efficient-VVC-Bitrate-Ladder-Prediction}, the authors created an ensemble aggregator to construct bitrate ladders, using the ensemble approach that contains a classifier that predicts the optimal video resolution at each bitrate, and a regressor that predicts cross-over bitrates between two resolutions. The authors again employed low-level features like GLCM and TC.

Instead of constructing content-gnostic bitrate ladders, the authors of \cite{Efficient-Per-Shot-Convex-Hull-Prediction-By-Recurrent-Learning} employed deep learning models that predict RQ points on the convex hull, by modeling the problem as a multi-label classification problem. A deep learning model employing Conv-GRU units was trained to predict RQ points on a convex hull using spatio-temporal features extracted from source videos. The authors also employed incremental learning to achieve tractable memory footprints and transfer learning to analyze wide ranges of content complexities, to augment the training of their deep learning models.

In our prior work \cite{Bitrate-Ladder-Construction-using-Visual-Information-Fidelity}, we deployed multiple Visual Information Fidelity (VIF) \cite{VIF} feature sets extracted over different scales and subbands of source videos to train regressors to predict the quality of compressed videos, given resolution and bitrate. We used these models to construct per-shot bitrate ladders. One advantage of using VIF feature sets is that they are readily available when using the widely deployed full-reference (FR) quality predictor VMAF. In \cite{Constructing-Per-Shot-Bitrate-Ladders-using-Visual-Information-Fidelity}, we studied the efficacy of using VIF features when constructing per-shot quality ladders along with bitrate ladders. We found that using an ensemble of content-aware low-level features and quality-aware VIF features to train regressors to predict the quality and bitrate, respectively, of compressed videos, resulted in superior performance as compared to employing either set of features individually.

\section{Method}
\label{sec:method}
\begin{figure}
   \centering
   \includegraphics[width=\columnwidth]{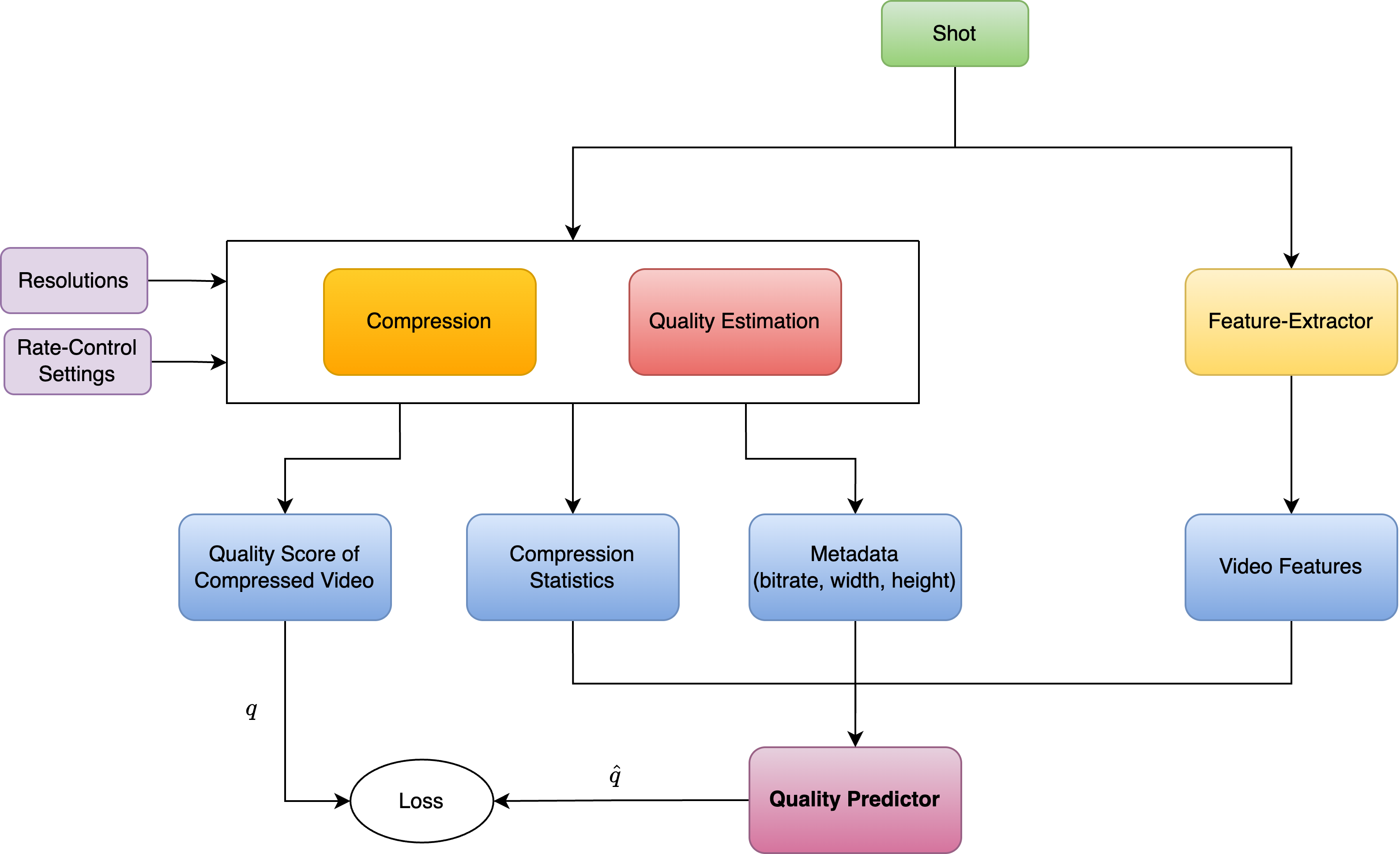}
   \caption{Block diagram demonstrating the training procedure of our proposed framework.}
   \label{fig:Framework:New-Proposed-Framework-Training}
\end{figure}

\begin{figure}
   \centering
   \includegraphics[width=0.75\columnwidth]{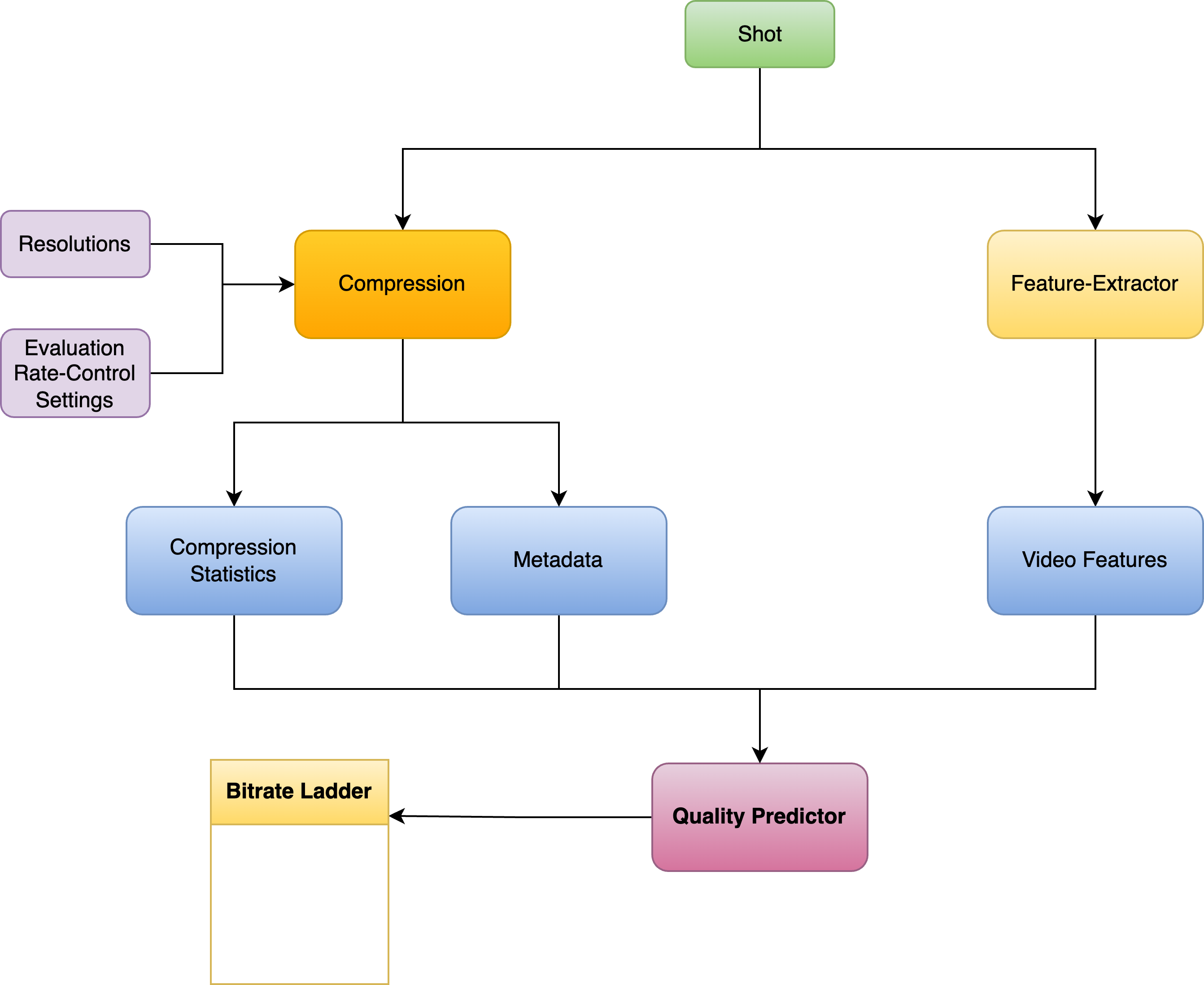}
   \caption{A block diagram demonstrating the inference procedure of our proposed bitrate ladder prediction framework.}
   \label{fig:Framework:New-Proposed-Framework-Testing}
\end{figure}

We propose a novel, transferable approach to constructing per-shot bitrate ladders that exploits variations induced by video compression, across various encoding settings. Figs. \ref{fig:Framework:New-Proposed-Framework-Training} and \ref{fig:Framework:New-Proposed-Framework-Testing} present the training and testing blocks of our proposed framework. The model we build goes significantly beyond our prior work \cite{Constructing-Per-Shot-Bitrate-Ladders-using-Visual-Information-Fidelity}, as we deploy highly accessible compression statistics to train our quality prediction regressors along with metadata, including the bitrates, widths, and heights of the compressed videos and perceptual video features. The inference procedure contains the compression module, unlike \cite{Constructing-Per-Shot-Bitrate-Ladders-using-Visual-Information-Fidelity}. During inferencing, video features are extracted from each source video shot or scene, which is compressed over all resolutions and a small number of rate-control settings to extract compression statistics. Instead of employing VMAF to estimate quality, we use a quality prediction regressor that is trained to predict the VMAF scores of compressed video.

We define compression statistics as (non-metadata) features that can be extracted from the compression process through logs, minor code modifications, or using \textit{ffprobe}. These features provide valuable insights into the changes incurred due to compression, eliminating the need for feature calculation on a compressed video when predicting quality. The (non-metadata) compression statistics that we use include the average bitrate and QP values of I, P, and B frames. When using H.265, these features are easily extracted from the compression logs. We leverage these compression statistics and the metadata when training regressors to predict the quality of compressed videos accurately, and subsequently to predict per-shot bitrate ladders having smaller losses against the convex hull.

We also extract video features from each source video to train the quality prediction regressor to be content-adaptive. We experimented with content-aware low-level features, quality-aware VIF features, and an ensemble of low-level content features and VIF features. Table \ref{table:features-sets} shows the sets of features considered in our experiments, while Table \ref{table:low-level-features} describes the low-level features considered, where $F_{1}$ denotes spatial feature pooling, $F_{2}$ denotes temporal feature pooling, $GLCM$ is the gray-level co-occurrence matrix \cite{GLCM}, $Coherence$ measures the spectral magnitude coherence between adjacent frames \cite{TC}, $CF$ is Colorfulness \cite{CF}, $Y$ is luminance, $U$ and $V$ are chroma components, $E_{Y}$ , $E_{U}$, $E_{V}$ are spatial energy components, $h_{Y}$ , $h_{U}$, $h_{V}$ are temporal energy components, $L_{Y}$ , $L_{U}$, $L_{V}$ are luminescence components, and $b$ is bitrate. We computed VIF features as follows:
\begin{align}
   \mathcal{C} = \mathcal{S} . \mathcal{U} = \{S_{i}.\vec{U}_{i}: i \in I\} \\
   \mathcal{E} = \mathcal{C} + \mathcal{N} \\
   I(\vec{C}^N;\vec{E}^N | s^N) = \sum_{j=1}^{N}\sum_{i=1}^{N} I(\vec{C}_{i};\vec{E}_{j} | \vec{C}^{i-1}, \vec{E}^{j-1}, s^N) \\
   I(\vec{C}^N;\vec{E}^N | S^N = s^N) = \sum_{i=1}^{N} I(\vec{C}_{i};\vec{E}_{i} | s_{i}) \\
   I(\vec{C}^N;\vec{E}^N | s^N) = \frac{1}{2}\sum_{i=1}^{N} \log_{2}(\frac{|s_{i}^2\mathbf{C_{U}} + \sigma_{n}^{2}\mathbf{I}|}{|\sigma_{n}^{2}\mathbf{I}|}), \label{eqn:vif0} \\
   I(\vec{C}^N;\vec{E}^N | s^N) = \frac{1}{2}\sum_{i=1}^{N}\sum_{j=1}^{M} \log_{2}(1 + \frac{s_{i}^2\lambda_{j}}{\sigma_{n}^{2}}) \\
   I_{k,b}^{j} = \frac{1}{N}\sum_{i=1}^{N} \log_{2}(1 + \frac{s_{i}^2\lambda_{j}}{\sigma_{n}^{2}}) .
\end{align}

Here $\mathcal{C} = \{\vec{C}_{i}: i \in I\}$ is a random field (RF) representing a subband of the reference image, $\mathcal{S} = \{\vec{S}_{i}: i \in I\}$ is a random field (RF) of positive scalars, and $\mathcal{U} = \{\vec{U}_{i}: i \in I\}$ is a Gaussian RF having mean zero and covariance $\mathbf{C}_{U}$. $\vec{C}_{i}$ and $\vec{U}_{i}$ are \textit{M}-dimensional vectors where $\vec{U}_{i}$ is independent of $\vec{U}_{j}$ when $i \neq j$. $\mathcal{E} = \{\vec{E}_{i}: i \in I\}$ is the output of the neural model and $\mathcal{N} = \{\vec{N}_{i}: i \in I\}$ models noise and uncertainty in the wavelet domain as a multivariate Gaussian of mean zero and covariance $\mathbf{C}_{N} = \sigma_{n}^2\mathbf{I}$. $I(\vec{C}^N;\vec{E}^N | s^N)$ represents the information that could ideally be extracted from a particular subband in the image. $\vec{s}^N$ is a realization of $S^N = (S_{1}, S_{2}, \dots, S_{N})$ on a particular reference image, which can be thought of as model parameters associated with it. We consider $I_{k,b}^{j}$ to represent mutual information along the $j^{th}$ eigenvector of the $b^{th}$ subband at the $k^{th}$ scale, where $j \in \{1,2,...,M\}$, $b \in \{1,2\}$, and $k \in \{1,2,3,4\}$. We fixed $M = 9$ in our experiments. We calculated the VIF feature set ($\text{VIFF}$) in Table \ref{table:features-sets} by temporally (mean) pooling the following features: $I_{k,b}^{j}$ along $F_{i}$ i.e luminance component of $i^{\text{th}}$ frame of a source video, $I_{k,b}^{j}$ along $D_{i} = F_{i} - F_{i-1}$, and the mean absolute luminance difference between consecutive frames. We compute these low-level features and the VIF features on each video, after resizing to dimensions $3840 \times 2160$.

\section{Dataset and Experimental Settings}
\label{sec:dataset-experimental-settings}
\subsection{Dataset}
We collected a comprehensive set of videos from the BVI-DVC dataset \cite{BVI-DVC} and the AV2 candidate test media from Xiph \cite{AV2}. We selected 175 videos from the BVI-DVC dataset including videos drawn from Videvo Free Stock Video Footage \cite{Videvo}, IRIS32 Free 4K Footage \cite{IRIS32}, the Mitch Martinez Free 4K Stock Footage \cite{MM}, the Dareful Free 4K Stock Video \cite{Dareful}, the TUM HD databases \cite{TUM}, the SJTU 4K video database \cite{SJTU}, the MCML 4K database \cite{MCML},  MCL-V \cite{MCL-V}, MCL-JCV \cite{MCL-JCV}, BVI Texture \cite{BVI-Texture}, BVI-HFR \cite{BVI-HFR}, LIVE-Netflix \cite{LIVE-Netflix-1, LIVE-Netflix-2}, Netflix Chimera \cite{Netflix-Chimera}, and the Ultra Video Group \cite{UVG}. We also selected 42 uncompressed AV2 4K resolution SDR test video sequences. The overall dataset of 217 videos comprises sequences of resolutions 3840x2160, 3840x2176, and 4096x2160, with frame rates ranging from 23 to 120 fps. The video sequences employ chroma sampling formats of either YUV420 or YUV422 and a bit depth of 10 bits per pixel. Each video in the dataset typically consists of a single scene without any scene cuts and was clipped to a maximum of 64 frames.

\subsection{Experimental Settings}
\begin{table}
   \centering
   \renewcommand{\arraystretch}{2}
   \caption{Feature sets.}
   \resizebox{\columnwidth}{!}{
   \begin{tabular}{| m{8em} | m{18em} | m{6em} |} 
   \hline
   \textbf{Notation for feature set} & \textbf{Features} & \textbf{No.of Features} \\ 
   \hline
   $\text{LLF}_{1}$ & GLCM, TC, SI, TI, CTI, CF, CI, DCT-Texture & 93 \\ 
   \hline
   $\text{LLF}_{2}$ & GLCM, TC, SI, TI, CTI, CF, CI, DCT-Texture, Bitrate-DCT-Texture & 96\\ 
   \hline
   $\text{VIFF}$ & $I_{k,b}^{j}[F_{i}]$, $|D_{i}|$, $I_{k,b}^{j}[D_{i}]$ & 145\\
   \hline
   \end{tabular}}
   \label{table:features-sets}
\end{table}

We employed \textit{ffmpeg} to perform compression and quality estimation. We calculated the quality scores of compressed videos using the popular full-reference video quality predictor VMAF \cite{VMAF}, which is available in ffmpeg. VMAF has been shown to exhibit higher correlations with human quality judgments than PSNR. We compute the VMAF score of each compressed video after upsampling it to a resolution of 3840$\times$2160 using Lanczos interpolation. Since we are interested in the transferability of per-shot bitrate ladders constructed across different encoder settings (codecs and presets), we compressed each of the 217 source videos using standard video codecs, including \textbf{libx265} (H.265), \textbf{libsvtav1} (ffmpeg with SVT-AV1 2.3.0), \textbf{libvpx-vp9} (VP9), and \textbf{libaom-av1} (AV1). We considered the presets `veryfast', `fast', `medium', and `slow' for \textbf{libx265}, `8', `6', and `4' for \textbf{libsvtav1}, `4' and `3' for \textbf{libvpx-vp9}, and `7' and `5' for \textbf{libaom-av1}. Our objective is to be able to construct bitrate ladders for fast encoder settings, then observe their performance when tested on slower encoder settings. We considered the following resolutions in our experiments: $\{3840 \times 2160, 2560 \times 1440, 1920 \times 1080, 1280 \times 720, 960 \times 540\}$.

\begin{figure*}
    \centering
    \begin{subfigure}[b]{0.24\textwidth}
        \centering
        \includegraphics[width=\textwidth]{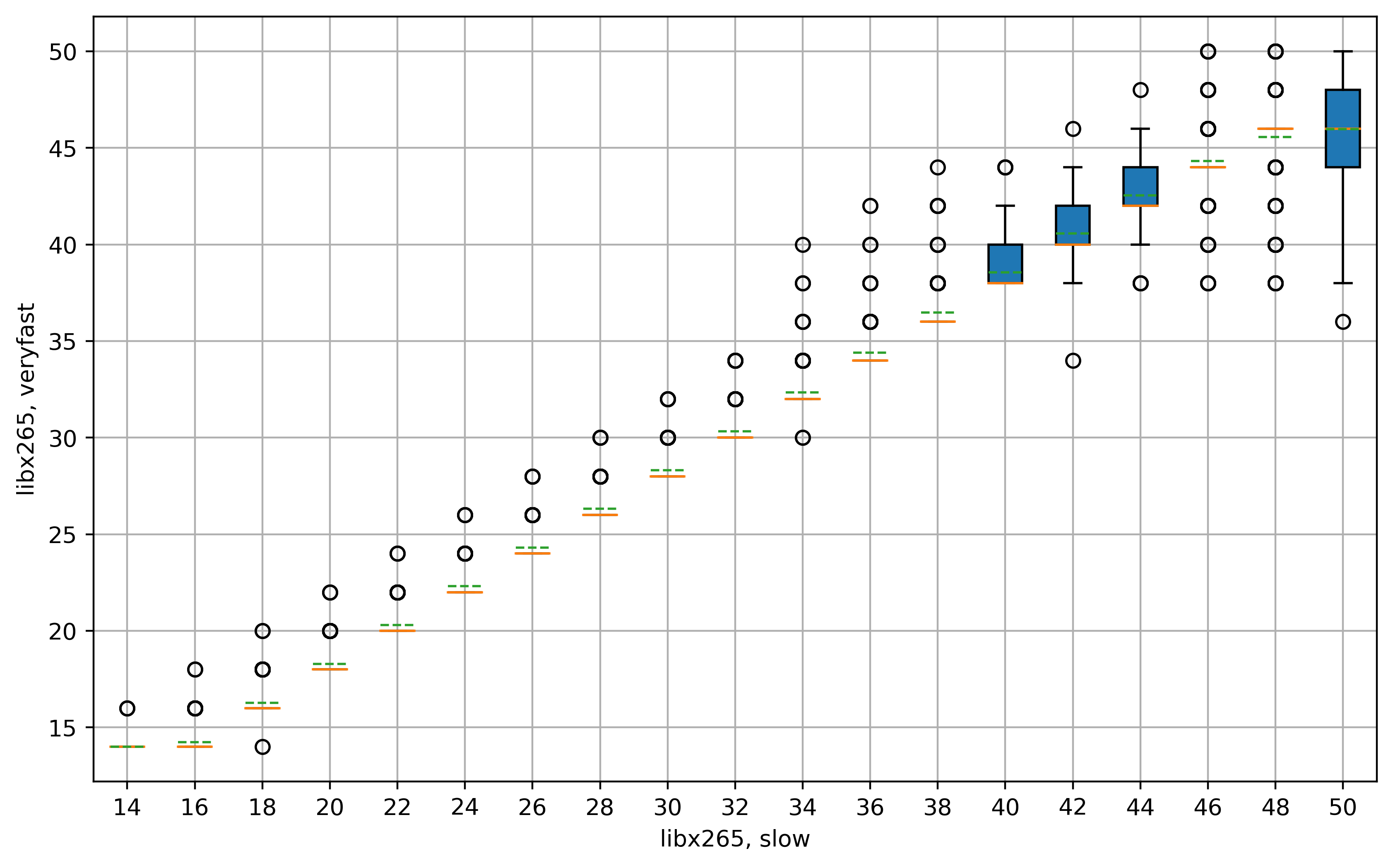}
        \caption{Mapping between CRFs of (libx265, slow) to (libx265, veryfast).}
        \label{fig:CRF_Map:libx265_veryfast-libx265_slow}
    \end{subfigure}
    \hfill
    \centering
    \begin{subfigure}[b]{0.24\textwidth}
        \centering
        \includegraphics[width=\textwidth]{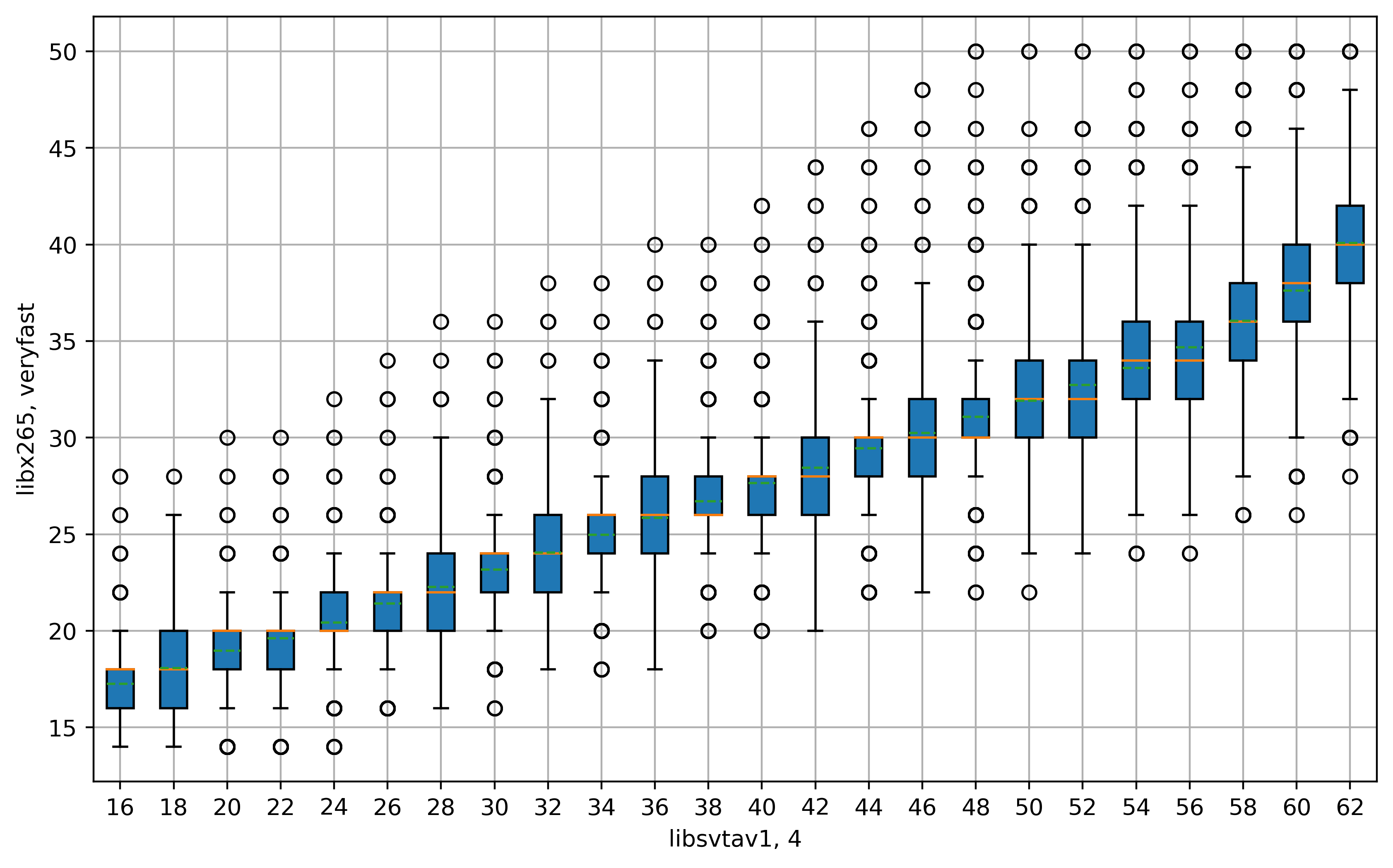}
        \caption{Mapping between CRFs of (libsvtav1, 4) to (libx265, veryfast).}
        \label{fig:CRF_Map:libx265_veryfast-libsvtav1_4}
    \end{subfigure}
    \hfill
    \begin{subfigure}[b]{0.24\textwidth}
        \centering
        \includegraphics[width=\textwidth]{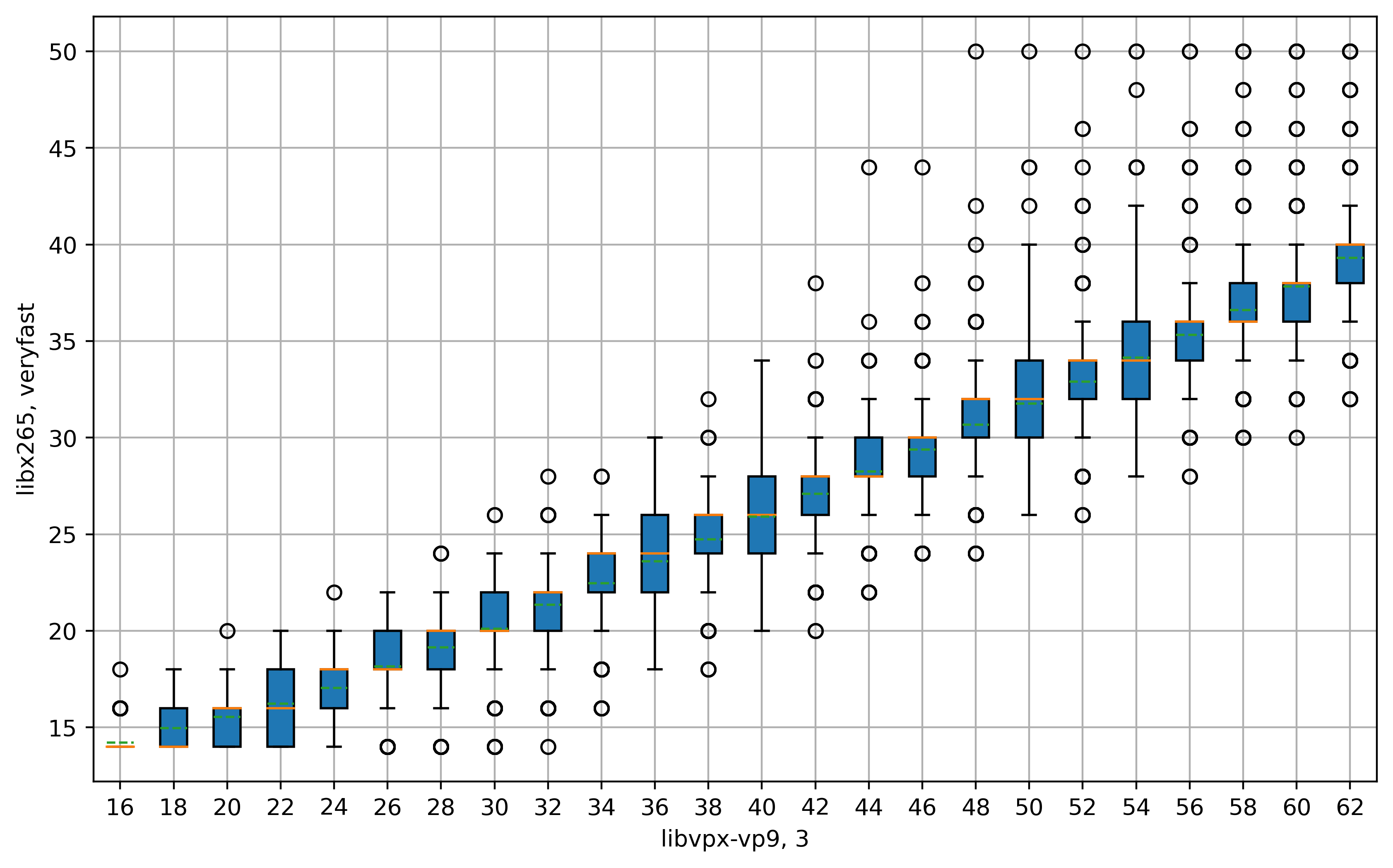}
        \caption{Mapping between CRFs of (libvpx-vp9, 3) to (libx265, veryfast).}
        \label{fig:CRF_Map:libx265_veryfast-libvpx-vp9_3}
    \end{subfigure}
    \hfill
    \begin{subfigure}[b]{0.24\textwidth}
        \centering
        \includegraphics[width=\textwidth]{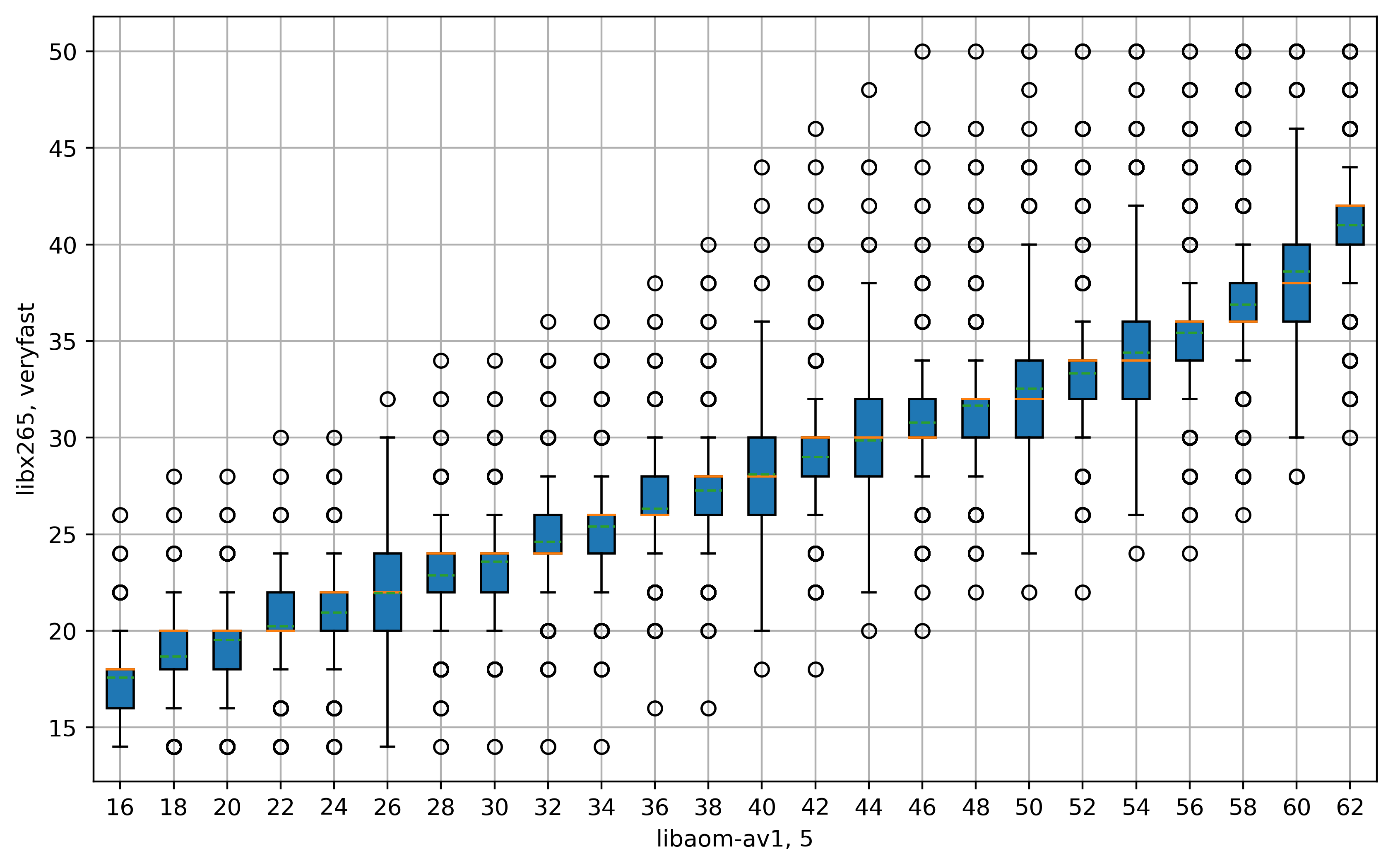}
        \caption{Mapping between CRFs of (libaom-av1, 5) to (libx265, veryfast).}
        \label{fig:CRF_Map:libx265_veryfast-libaom-av1_5}
    \end{subfigure}
    \caption{Mapping between CRFs from different (codec, preset) settings to (libx265,veryfast).}
    \label{fig:CRF_Map}
\end{figure*}

We applied different CRF ranges depending on the different encoders used. For \textbf{libx265} we varied the CRFs from 14 to 50 (inclusive) with a skip of 2, and for \textbf{libsvtav1}, \textbf{libvpx-vp9}, and \textbf{libaom-av1}, we varied the CRF values from 16 to 62 (inclusive) with a skip of 2. We sampled CRFs based on CRF maps we constructed between different encoder settings. Figure \ref{fig:CRF_Map} shows the distribution of optimal CRFs of (libx265, veryfast) encoder settings, where for each CRF (codec, preset) setting, we identified those CRF (libx265, veryfast) settings that produced the best matching bitrate. It may be observed that for the slow preset of libx265, the mapping between CRFs is nearly linear, with increased deviations at high CRF values. By contrast, for (libsvtav1, “4”), (libvpx-vp9, 3), and (libaom-av1, 5), which share the same CRF ranges, the higher CRFs of these codecs yielded relatively smaller optimal CRFs when using the (libx265, veryfast) encoder settings. This shows that, as compared to the (libx265, veryfast) encoder settings, the considered set of encoder settings generally yield higher bitrates (smaller CRFs of libx265). However, it may also be observed that there are a substantial number of outliers for each distribution of optimal CRFs. Since the sample size of 217 videos is insufficient to draw definitive conclusions regarding CRF mappings between encoder settings, in our experiments, we used the above-mentioned CRF ranges to cover all scenarios.

\begin{table}
   \renewcommand{\arraystretch}{2}
   \normalfont
   \normalsfcodes
   \centering
   \caption{List of low-level features.}
   \resizebox{\columnwidth}{!}{
   \begin{tabular}[]{| m{4em} | m{30em} | m{4em} |}
        \hline
        \textbf{Feature} & \textbf{Formula} & \textbf{No.of Features}\\
        \hline
        GLCM & $F_{2}\{F_{1}\{correlation(GLCM)\}\}$, $F_{2}\{F_{1}\{contrast(GLCM)\}\}$, $F_{2}\{F_{1}\{energy(GLCM)\}\}$, $F_{2}\{F_{1}\{homogeneity(GLCM)\}\}$ where GLCM is calculated on blocks of size (64,64), $F_{1} = \{mean, std\}$ and $F_{2} = \{mean, std, skew, kurtosis\}$ & 32\\
        \hline
        TC & $F_{2}\{F_{1}\{Coherence\}\}$ where $F_{1} = \{mean, std, skew, kurtosis\}$ and $F_{2} = \{mean, std\}$ & 8\\
        \hline
        SI & $F_{2}\{F_{1}\{Sobel(Y)\}\}$ where $F_{1} = \{mean, std\}$ and $F_{2} = \{mean, std, skew, kurtosis\}$ & 8\\
        \hline
        TI & $F_{2}\{F_{1}\{(Y_{i+1} - Y_{i})\}\}$ where $F_{1} = \{mean, std\}$ and $F_{2} = \{mean, std, skew, kurtosis\}$ & 8\\
        \hline
        CTI & $F_{2}\{F_{1}\{Y\}\}$ where $F_{1} = \{mean, std\}$ and $F_{2} = \{mean, std, skew, kurtosis\}$ & 8\\
        \hline
        CF & $F_{2}\{CF(YUV)\}$ where $F_{2} = \{mean, std, skew, kurtosis\}$ & 4\\
        \hline
        CI & $F_{2}\{F_{1}\{U\}\}$, $F_{2}\{W_{R} \times F_{1}\{V\}\}$ where $W_{R} = 5$, $F_{1} = \{mean, std\}$ and $F_{2} = \{mean, std, skew, kurtosis\}$ & 16\\
        \hline
        DCT-Texture & $F_{2}\{E_{Y}\}$, $F_{2}\{h_{Y}\}$, $F_{2}\{L_{Y}\}$, $F_{2}\{E_{U}\}$, $F_{2}\{h_{U}\}$, $F_{2}\{L_{U}\}$, $F_{2}\{E_{V}\}$, $F_{2}\{h_{V}\}$, $F_{2}\{L_{V}\}$ where $F_{2} = \{mean\}$ & 9\\
        \hline
        Bitrate-DCT-Texture & $\log_{2}\left[\sqrt{\frac{F_{2}\{h_{Y}\}}{F_{2}\{E_{Y}\}}}\right] + 2\log_{2}(b)$, $\log_{2}\left[\sqrt{\frac{F_{2}\{h_{U}\}}{F_{2}\{E_{U}\}}}\right] + 2\log_{2}(b)$, $\log_{2}\left[\sqrt{\frac{F_{2}\{h_{V}\}}{F_{2}\{E_{V}\}}}\right] + 2\log_{2}(b)$ where $F_{2} = \{mean\}$ & 3\\
        \hline
   \end{tabular}}
   \label{table:low-level-features}
\end{table}

\begin{figure*}
    \centering
    \begin{subfigure}[b]{0.24\textwidth}
        \centering
        \includegraphics[width=\textwidth]{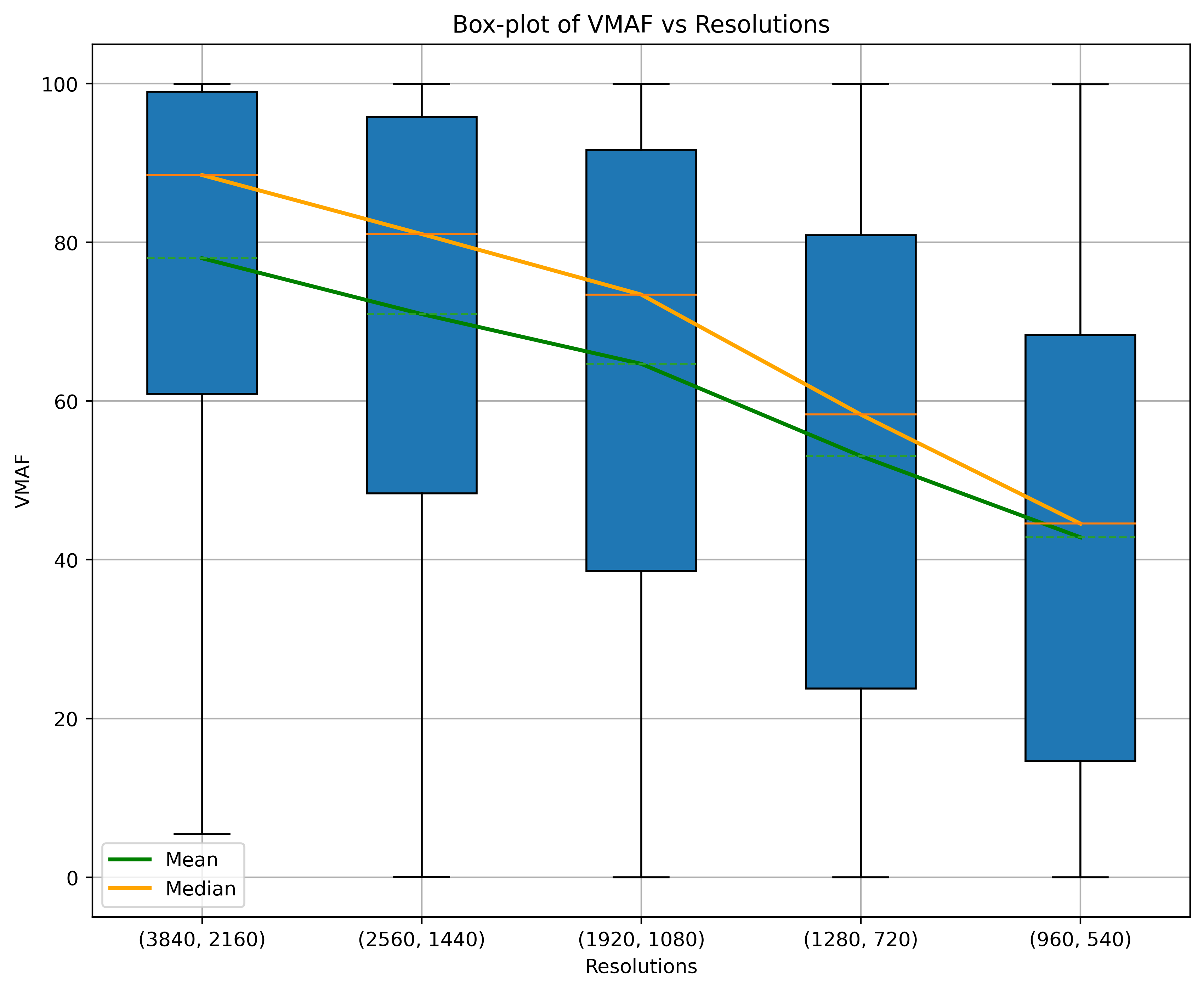}
        \caption{(libx265, slow)}
        \label{fig:box_plots_VMAF:libx265_slow}
    \end{subfigure}
    \hfill
    \begin{subfigure}[b]{0.24\textwidth}
        \centering
        \includegraphics[width=\textwidth]{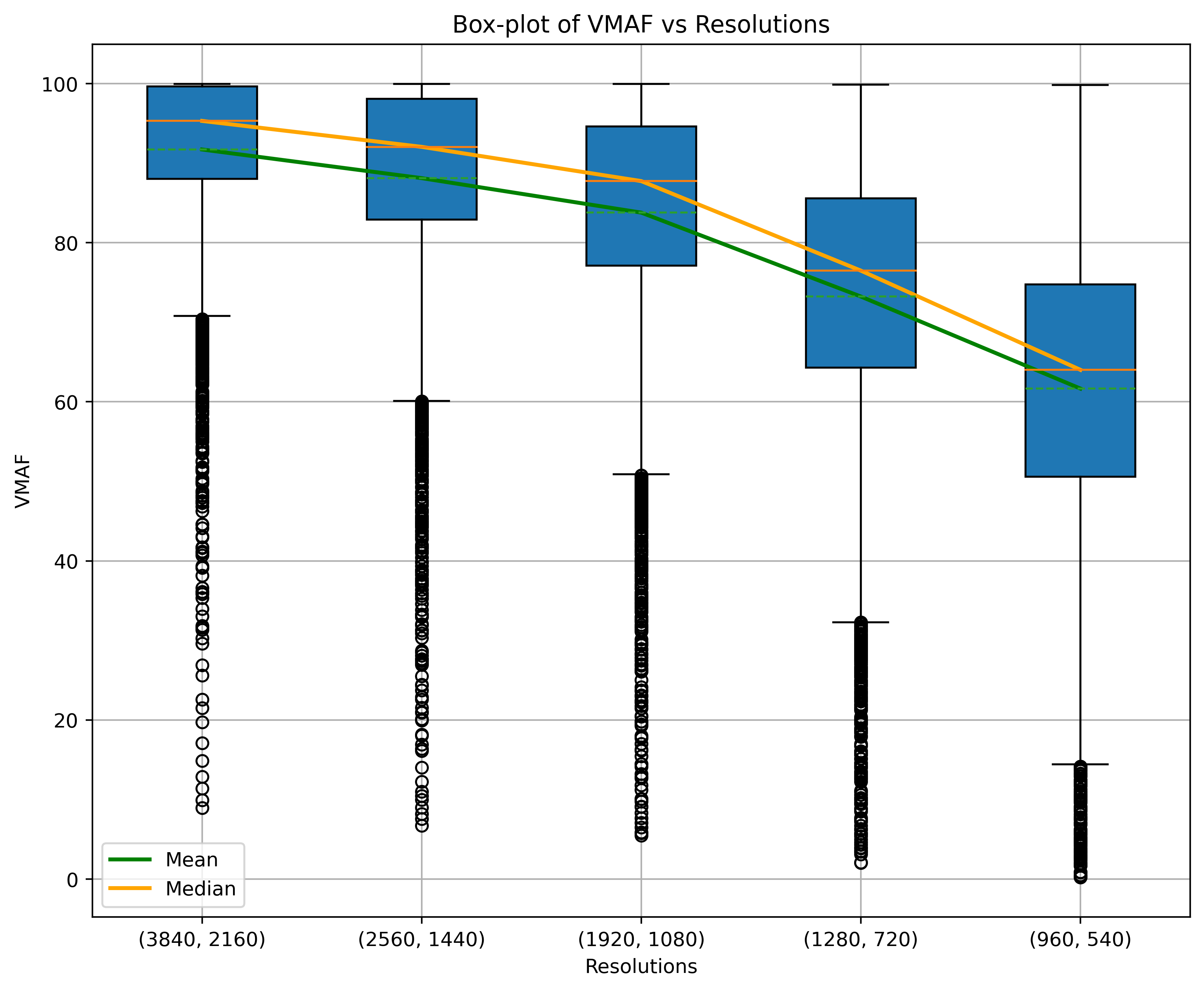}
        \caption{(libsvtav1, 4)}
        \label{fig:box_plots_VMAF:libsvtav1_4}
    \end{subfigure}
    \hfill
    \begin{subfigure}[b]{0.24\textwidth}
        \centering
        \includegraphics[width=\textwidth]{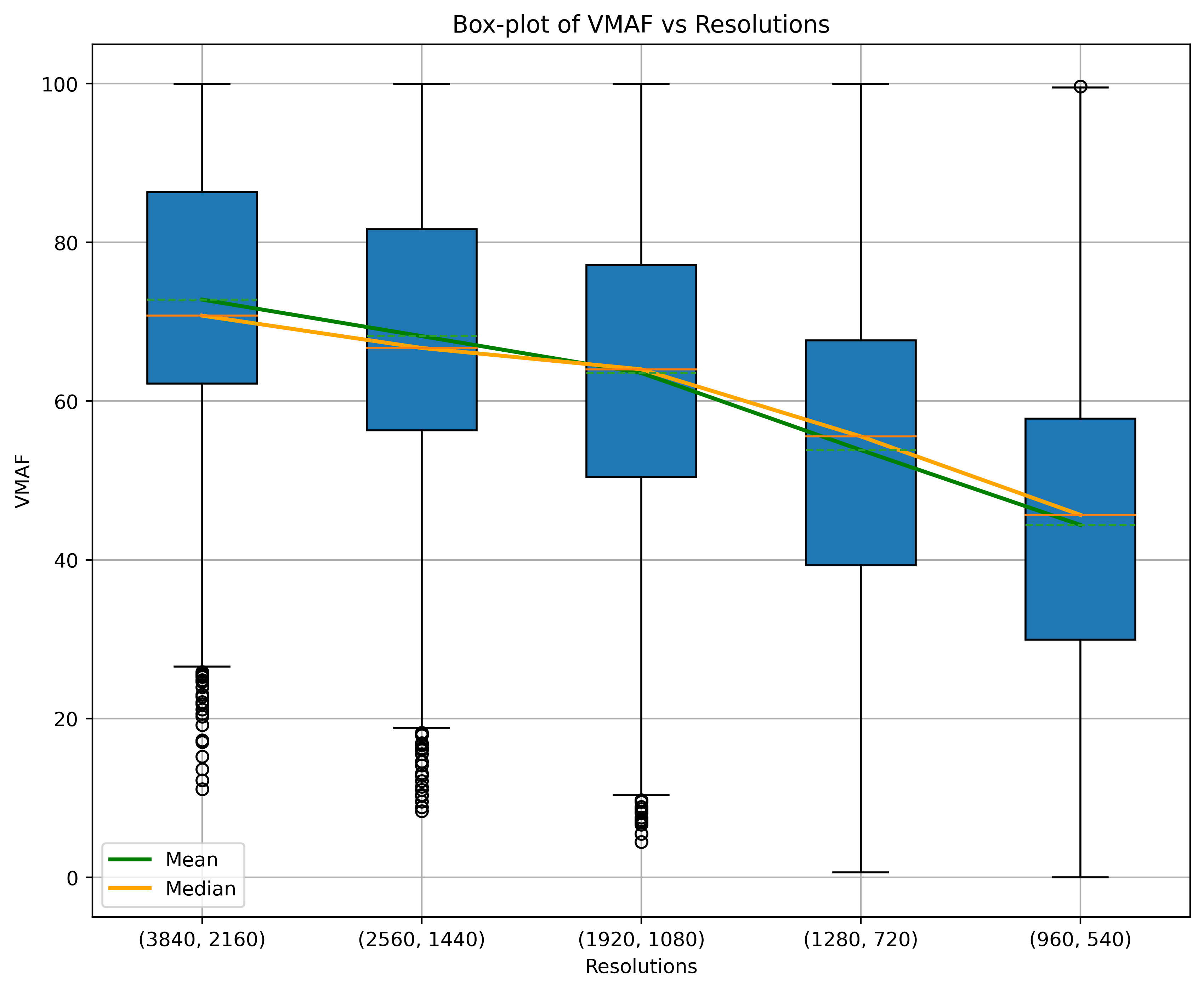}
        \caption{(libvpx-vp9, 3)}
        \label{fig:box_plots_VMAF:libvpx-vp9_3}
    \end{subfigure}
    \hfill
    \begin{subfigure}[b]{0.24\textwidth}
        \centering
        \includegraphics[width=\textwidth]{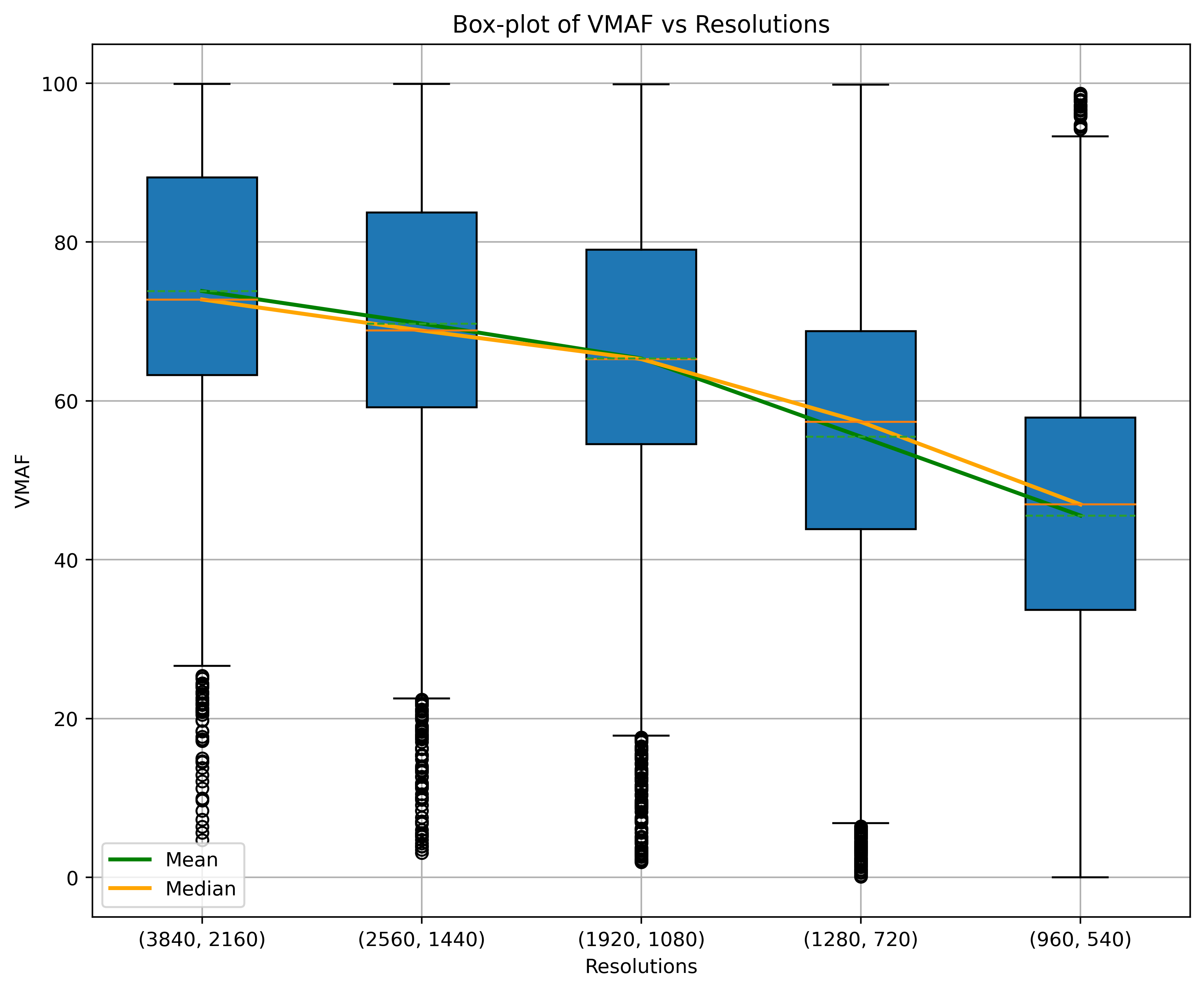}
        \caption{(libaom-av1, 5)}
        \label{fig:box_plots_VMAF:libaom-av1_5}
    \end{subfigure}
    \caption{Distribution of VMAF scores at each considered resolution for various encoder settings.}
    \label{fig:box_plots_VMAF}
\end{figure*}

To determine the quality constraints, we examined the distributions of VMAF scores at each resolution, for the slowest presets for each codec in our considered encoder settings. Fig. \ref{fig:box_plots_VMAF} shows the distribution of VMAF scores at each resolution over all the videos in the dataset and CRFs considered in each codec. It may be observed that the libx265 codec with a slow preset yielded a wide range of VMAF scores at each resolution with a good interquartile range. Figure \ref{fig:box_plots_VMAF:libsvtav1_4} shows that videos compressed using the libsvtav1 codec using preset 4 led to very high VMAF scores across resolutions, but accompanied by a substantial number of outliers as compared to other encoder settings. In general, at high resolutions, the VMAF scores of videos compressed using (libx265, slow) and (libsvtav1, 4) delivered higher median VMAF scores as compared to the encoder settings (libvpx-vp9, 3) and (libaom-av1, 5). For the majority of the considered encoder settings, a substantial portion of the compressed videos had VMAF scores $\geq$ 20. Since lower VMAF scores are associated with unsatisfactory visual experiences, we constrained the quality range to (20, 99.9). We fixed the upper limit of 99.9 to prevent training on RQ points saturated at 100. We used the following experimental settings throughout:
\begin{itemize}
    \item \textbf{(Codec, Presets):} (libx265, \{veryfast, fast, medium, slow\}), (libsvtav1, \{8, 6, 4\}), (libvpx-vp9, \{4, 3\}), and (libaom-av1, \{7, 5\})
    \item \textbf{Resolutions:} 3840$\times$2160, 2560$\times$1440, 1920$\times$1080, 1280$\times$720, 960$\times$540
    \item \textbf{(Codec, CRFs):} (libx265, \{14,16,18,...,50\}), (libsvtav1, \{16,18,20,...,62\}), (libvpx-vp9, \{16,18,20,...,62\}), and (libaom-av1, \{16,18,20,...,62\})
    \item \textbf{Constraints}: 20 $\leq$ VMAF $\leq$ 99.9
\end{itemize}

\section{Experiments}
\label{sec:experiments}
Next, we present the experiments conducted using our proposed framework, prior leading methods, a complexity analysis, and the evaluation metrics employed to evaluate performance.

\subsection{Proposed Method}
\subsubsection{Training}
As described in Section \ref{sec:method}, our model employs compression statistics along with features derived from the source video and the metadata of the compressed video to train regressors to predict the quality of compressed videos. In the experiments, we extracted the average QP and bitrate of I, P, and B frames as compression statistics. We trained four different regressors employing four different sets of video features, including no video features, content-aware low-level features ($\text{LLF}_{2}$), quality-aware VIF features ($\text{VIFF}$), and an ensemble of low-level features and VIF features ($\text{LLF}_{2}$, $\text{VIFF}$), along with metadata and compression statistics to predict the quality of compressed videos. We trained an Extra-Trees regressor to predict the quality of compressed videos based on its successes in prior works \cite{Constructing-Per-Shot-Bitrate-Ladders-using-Visual-Information-Fidelity, Bitrate-Ladder-Construction-using-Visual-Information-Fidelity, Benchmarking-Learning-based-Bitrate-Ladder-Prediction-Methods-for-Adaptive-Video-Streaming}.
\subsubsection{Inference}
During inference, our approach requires compressing each source video to extract compression statistics, which are subsequently employed to predict the quality of the compressed video. Consequently, we restricted the number of times compression was applied, in view of the increase in time complexity. Hence, during inference, we compressed each source video across all resolutions using CRFs \{18,22,26,30,34,38,42\} if the libx265 codec was employed as a fast encoder, or \{20,26,32,38,44,50,56,62\} if libsvtav1, libvpx-vp9, or libaom-av1 codecs were utilized as fast encoders. The quality prediction regressors were then used to predict the quality of the compressed videos, rather than using VMAF, during the construction of per-shot bitrate ladders. At each step of the bitrate ladder, based on interpolated predicted quality scores, the resolution yielding the best quality score (at that bitrate) was selected as the optimal resolution.

\subsection{Methods for Comparison}
We compared the performance of our model against standard and commonly used bitrate ladder construction and video encoding techniques used by industry, and ML-based methods. We shall describe them as follows:

\subsubsection{Fixed Bitrate Ladder}
We evaluated the performance of various bitrate ladder construction methods against the fixed bitrate ladder proposed by Apple \cite{Fixed-Bitrate-Ladder}. Although the fixed bitrate ladder is primarily designed for the libx265 codec, we employed the same fixed bitrate ladder on other codecs when evaluating the transferability of constructed per-shot bitrate ladders. However, since we are applying the fixed bitrate ladder designed for libx265 to other codec settings, in certain scenarios, no RQ points were available to construct an RQ curve. Consequently, we considered the gains relative to the fixed bitrate ladder to be zero, which is a very conservative approach.

\subsubsection{Convex Hull}
Similar to the fixed bitrate ladder, we evaluated the performance of the compared methods against the convex hull created by exhaustively encoding a video across resolutions and CRFs. We constructed a unique convex hull for each video across all different encoding settings, i.e., codec and preset pairs.

\subsubsection{Two-Step Convex-Hull}
As described in \cite{Fast-Encoding-Parameter-Selection-for-Convex-Hull-Video-Encoding}, we constructed the convex hull only for fast encoder settings on each video in the dataset. During inference, we employed the same convex hull for all other encoding settings without any modification.

\subsubsection{Cross-Over Bitrates}
As described in \cite{Benchmarking-Learning-based-Bitrate-Ladder-Prediction-Methods-for-Adaptive-Video-Streaming}, we trained a regressor to predict the cross-over bitrate between adjacent resolutions using low-level video features ($\text{LLF}_{1}$) extracted from the source video. To mitigate the curse of dimensionality, we employed recursive feature elimination (RFE) to select a subset of nine features when training the regressor. We then used the predicted cross-over bitrates between higher resolutions as additional features, to predict cross-over points between lower resolutions. Finally, based on the predicted cross-over bitrates, we selected the optimal resolutions at bitrate steps and constructed per-shot bitrate ladders.

\subsection{Complexity}
\begin{figure}
    \centering
    \includegraphics[width=\columnwidth]{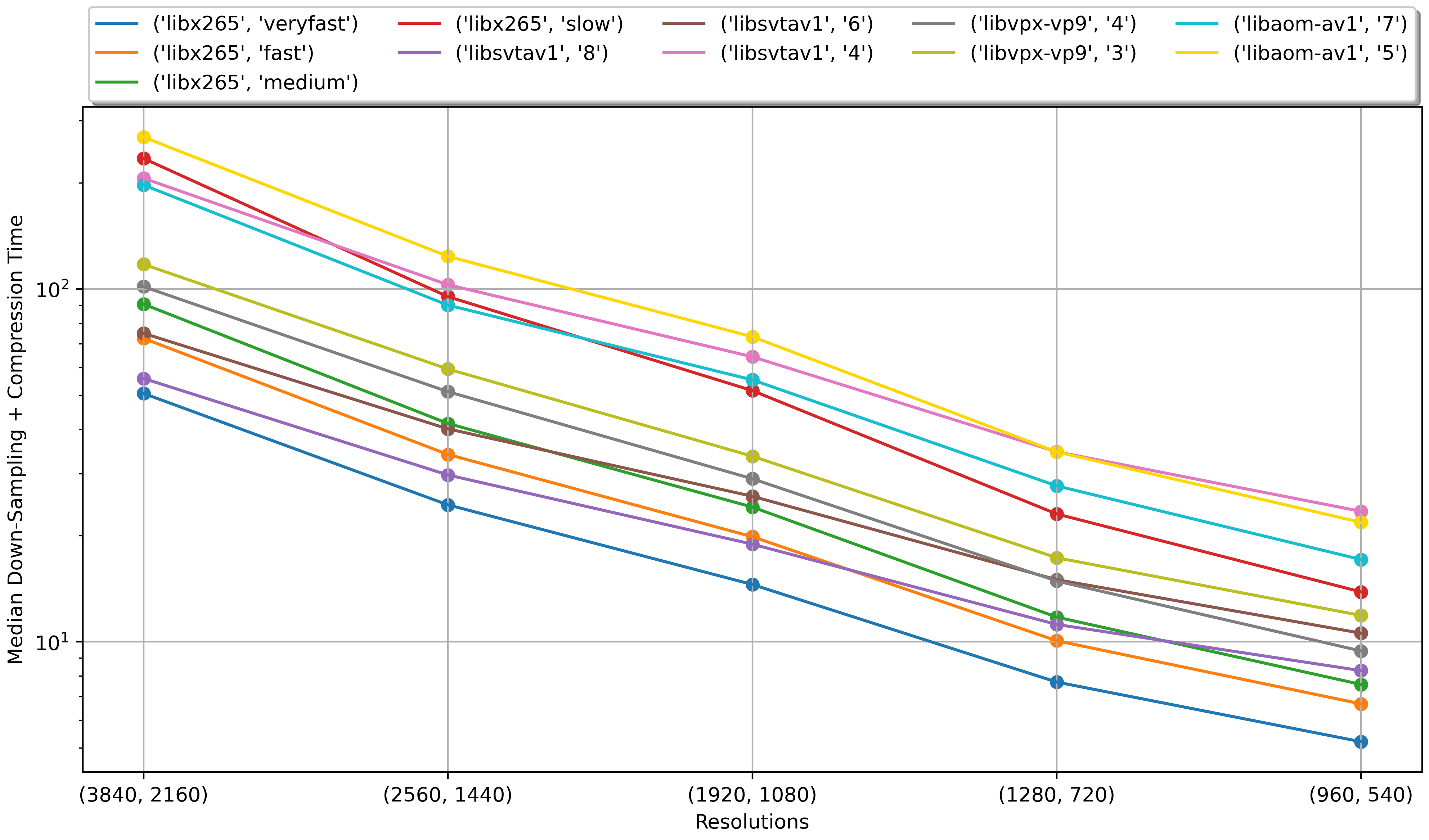}
    \caption{Median execution time (system + user) for downsampling and compression of videos in the dataset.}
    \label{fig:time-complexity}
\end{figure}

As described in the Introduction, we aim to construct per-shot bitrate ladders on a fast encoder, then evaluate their performances across relatively slow encoding settings. In our experiments, we utilized an AMD Ryzen Threadripper 3970X and reported the total execution time, which includes both system and user time. It is also worth noting that, the total execution time is different from the time elapsed. The execution time of a process varies based on CPU scheduling, system load, and other running processes. Hence, the times discussed here should be understood as providing a baseline understanding of the time complexities of each process, but do not necessarily represent the best achievable execution time.

We analyzed the time complexity of downsampling and video compression for each of the considered encoder settings at each resolution, for all videos and CRFs considered in our experiments. The execution times of the downsampling and compression processes vary with resolution, CRF, preset, and codec. We present the median execution time for each resolution and encoding setting (codec and preset). Fig. \ref{fig:time-complexity} plots the median downsampling and compression execution times (in seconds) across resolutions, where the y-axis is a log scale. It may be observed that the encoder setting (libx265, veryfast) required the least execution time, while (libx265, slow), (libsvtav1, 4), (libaom-av1, 7), and (libaom-av1, 5) needed lengthier execution times. As anticipated, the execution time decreased as the preset speed was increased or as the resolution was decreased. Considering these findings, we selected (libx265, veryfast) as the fast encoder. Table \ref{table:execution-times} lists the approximate median execution time of each process for all videos in the dataset. We observed that the median execution time for upsampling and calculating VMAF was in the range of 0.75 minutes. For a video with 64 frames and resolution $3840 \times 2160$, we observed that the execution time to calculate low-level features ($\text{LLF}_{1}$) was approximately 7.3 minutes, while the execution time to calculate VIF features ($\text{VIFF}$) was approximately 67 minutes. The disparity in the execution time between the calculation of VIF features and VMAF arises from the C-based code employed by \textit{ffmpeg} and the Python code used in our experiments. Estimating VMAF between a reference video and a compressed video involves calculating VIF features on both videos. However, our models only calculate VIF features on the reference video, which is the source video. Additionally, we observed an approximate execution time of 0.167 minutes for the Extra-Trees Regressor to predict quality, given the ensemble of video features as input.

We analyzed the time complexities of our proposed models and those considered for comparison. Table \ref{table:time_complexity} lists the time complexities of the compared methods. A simple lookup table, such as the fixed bitrate ladder \cite{Fixed-Bitrate-Ladder} requires no computation and is the same for all videos. However, constructing the convex hull would necessitate computing the video compression and VMAF modules approximately 95 to 120 times (5 resolutions $\times$ 19 or 24 CRFs) on each video and every encoding setting under consideration. The two-step convex hull method significantly reduces the complexity by constructing the convex hull only on a fast encoder, requiring only $95 \times t_{\text{libx265,veryfast}}$ for each video, and which can be applied to all other encoding techniques. Existing ML-based bitrate ladder construction methods that employ cross-over bitrates \cite{Benchmarking-Learning-based-Bitrate-Ladder-Prediction-Methods-for-Adaptive-Video-Streaming} do not employ any compression or quality estimation modules, instead relying on a feature extraction module that executes only once on the source video. Our proposed methods incorporate both feature extraction modules and features from compression using the fast encoder. The feature extraction module is only executed once, i.e., on the source video. However, it requires an additional $35 \times t_{\text{libx265,veryfast}}$ time, as compared to prior ML-based methods.

\begin{table}
    \normalfont
    \normalsfcodes
    \renewcommand{\arraystretch}{1.75}
    \centering
    \caption{Approximate execution times of the various processes involved in the construction of the per-shot ladders using various methods.}
    \label{table:execution-times}
        \resizebox{\columnwidth}{!}{
        \begin{tabular}{| m{12em} | m{8em} | m{11em} |}
        \hline
        \textbf{Process} & \textbf{Language} & \textbf{Median Execution Time} (in minutes) \\
        \hline
        Upsampling and VMAF & C & 0.75 \\
        \hline
        $\text{LLF}_{1}$ & Python & 7.3 \\
        \hline
        $\text{VIFF}$ & Python & 67 \\
        \hline
        Extra-Trees Regressor & Python & 0.167 \\
        \hline
    \end{tabular}}
\end{table}

\begin{table}
    \renewcommand{\arraystretch}{1.75}
    \normalfont
    \normalsfcodes
    \centering
    \tabcolsep=0.25cm
    \caption{Time-complexity analysis of compared methods}
    \resizebox{\columnwidth}{!}{
    \begin{tabular}{ c| c | c | c}
        \hline
        \shortstack[c]{{}\\\textbf{Method}\\{}} & \shortstack[c]{{}\\\textbf{Downsampling} \&\\\textbf{Compression}} & \shortstack[c]{{}\\\textbf{Upsampling} \&\\\textbf{Quality Estimation}} & \shortstack[c]{{}\\\textbf{Compute}\\\textbf{Features}} \\

        \hline

        Fixed Bitrate Ladder & 0 & 0 & No\\

        \hline

        Convex-Hull & (95 or 120) $ \times t_{\text{codec,preset}}$ & (95 or 120) $ \times t_{\text{VMAF}}$ & No\\

        Two-Step Convex-Hull & $95 \times t_{\text{libx265,veryfast}}$ & 95 $ \times t_{\text{VMAF}}$ & No\\

        \hline

        Cross-Over Bitrates & 0 & 0 & Yes\\

        \hline

        Our Proposed Method & $35 \times t_{\text{libx265,veryfast}}$ & 0 & Yes\\

        \hline
    \end{tabular}}
    \label{table:time_complexity}
\end{table}

\subsection{Evaluation Metrics}
We evaluated the compared methods on a large sample of videos, employing k-fold cross-validation for training, validation, and testing. We fixed $k=5$, hence each fold consisted of 20\% of the dataset. We split the 4 folds into training (90\%) and validation datasets (10\%), while the remaining fold was used for testing. We repeated this process, rotating through all folds to ensure each fold was tested once. We report average performance calculated based on predictions on all the test folds. We constructed per-shot bitrate ladders at bitrates (in kbps) \{100, 200, 400, 600, 800, 1000, 1500, 2000, 2400, 3000, 3500, 4000, 4500, 5000, 6000, 7000, 8100, 9000, 10000, 11600, 13000, 15000\} for each method under consideration. We considered the exhaustive set of bitrates mentioned earlier, as we conducted experiments across various codecs and preset settings, where each encoder setting resulted in a compressed video having a different bitrate. When employing the fixed bitrate ladder to construct the RQ curve, we used the smallest resolution, 540p, for the lowest bitrates. We also employed the Top-Bottom bitrate ladder correction method described in \cite{Constructing-Per-Shot-Bitrate-Ladders-using-Visual-Information-Fidelity}, to ensure monotonic changes in resolution in the predicted bitrate ladder.

We employed the Bjøntegaard Delta (BD) method \cite{BD} to evaluate the performances of our proposed methods against the fixed bitrate ladder \cite{Fixed-Bitrate-Ladder} and the convex hull constructed using exhaustive encoding. It measures the average difference between two RQ curves by fitting a third-order cubic polynomial through the data points, then calculating the integral to estimate the average bitrate savings, referred to as BD-Rate, and the average quality savings, referred to as BD-Quality (e.g., BD-PSNR or BD-VMAF). We employed the implementation in \cite{BD-Metric}, based on the study \cite{BD-Study}. Since we are performing constant-quality encoding, to construct the rate-quality curve using a bitrate ladder, we utilized the RQ points of the video from the dataset having resolution $R_{i}$, with bitrates lying in the range $[b_{i}, b_{i+1}]$, where $b_{i}$ is a step in the bitrate ladder and $R_{i}$ is its corresponding optimal resolution. We calculated the BD-Rate and BD-VMAF on each video in the test datasets against the fixed bitrate ladder and the convex hull. We report the mean BD-rate and BD-VMAF calculated across all the folds in the dataset, against the fixed bitrate ladder and the convex hull (instead of the reference bitrate ladder sampled from the convex hull). The negative values of BD-rate denote bitrate savings, while positive values of BD-Quality denote quality gains. The BD-rate is reported in percentage, while BD-VMAF is reported in VMAF units. To better understand the distribution of BD metrics, we report the closeness of the method to the convex hull using the BD metrics against the fixed bitrate ladder. We calculated the closeness of a method by estimating the fractions of samples yielding both BD-Rate savings and BD-VMAF gain greater than 75\% of the BD metrics of the convex hull against the fixed bitrate ladder, which we denote as $f_{75}$. Among all the experiments we conducted, including evaluations of various prior models, we observed that in less than 2.5\% of cases, we were unable to calculate the BD-rate/BD-VMAF against the fixed bitrate ladder due to a zero overlap between VMAF scores or bitrates, respectively.

\section{Results}
\label{sec:results}
Next, we discuss the performance of our proposed methods against prior methods, the fixed bitrate ladder \cite{Fixed-Bitrate-Ladder}, and the convex hull constructed using exhaustive encoding. As discussed earlier, we trained regressors to predict the quality of compressed videos, then employed these regressors to predict the per-shot bitrate ladder. We trained the regressors and constructed per-shot bitrate ladders using rate-quality points obtained by compressing the source videos using the decided fast encoder settings, \textit{viz}., the libx265 codec with veryfast preset. We evaluated the transferability of the obtained per-shot bitrate ladders across various other encoder settings.

\begin{table}
    \centering
    \tabcolsep=0.2cm
    \renewcommand{\arraystretch}{1.75}
    \resizebox{\columnwidth}{!}{
    \begin{tabular}{c c c c c c }
    Video-Features & (3840,2160) & (2560,1440) & (1920,1080) & (1280,720) & (960,540) \\
    \hline
    No features & 0.923 & 0.928 & 0.915 & 0.863 & 0.763 \\
    \hline
    $\text{LLF}_{2}$ & 0.958 & 0.953 & 0.946 & 0.919 & 0.888 \\
    \hline
    $\text{VIFF}$ & 0.957 & 0.952 & 0.945 & 0.925 & 0.901 \\
    \hline
    $\text{LLF}_{2}, \text{VIFF}$ & 0.965 & 0.962 & 0.957 & 0.94 & 0.917 \\
    \hline
    \end{tabular}}
    \caption{The average PLCC between true VMAF scores and VMAF scores predicted by Extra-Trees regressors trained on various video features, metadata, and compression statistics, across all dataset splits.}
    \label{table:libx265-PLCC_1}
\end{table}

\subsection{Effectiveness of Video Features}
\begin{table}
    \centering
    \tabcolsep=0.2cm
    \renewcommand{\arraystretch}{1.75}
    \resizebox{\columnwidth}{!}{
    \begin{tabular}{c c c c c c }
    Video-Features & (3840,2160) & (2560,1440) & (1920,1080) & (1280,720) & (960,540) \\
    \hline
    No features & 0.724 & 0.73 & 0.71 & 0.639 & 0.528 \\
    \hline
    $\text{LLF}_{2}$ & 0.824 & 0.825 & 0.81 & 0.766 & 0.701 \\
    \hline
    $\text{VIFF}$ & 0.799 & 0.804 & 0.795 & 0.766 & 0.73 \\
    \hline
    $\text{LLF}_{2}, \text{VIFF}$ & 0.839 & 0.842 & 0.832 & 0.799 & 0.754 \\
    \hline
    \end{tabular}}
    \caption{The average PLCC between true VMAF scores and VMAF scores predicted by Extra-Trees regressors trained on various video features and metadata, across all dataset splits.}
    \label{table:libx265-PLCC_2}
\end{table}

To demonstrate the effectiveness of the various sets of video features extracted from source videos, we evaluated the performance of the Extra-Trees regressors trained to predict the quality of compressed videos on each spatial resolution, over all the splits in the dataset. Table \ref{table:libx265-PLCC_1} shows the average PLCC values calculated between the true VMAF scores of compressed videos and the VMAF scores predicted by our proposed quality prediction regressors: Extra-Trees regressors trained on the video features, metadata, and compression statistics across all dataset splits. Similar to Table \ref{table:libx265-PLCC_1}, Table \ref{table:libx265-PLCC_2} shows the average PLCC values obtained by our proposed models when compression statistics were not employed: regressors trained only on video features and metadata. It may be observed that employing video features enhanced the performance of the quality prediction regressors as compared to employing only metadata. Among video features, the low-level features provided higher correlations on larger resolution videos, while employing VIF features yielded higher correlations on lower resolution videos. However, employing the ensemble of both low-level features and VIF features delivered even higher correlations between the true VMAF scores and the predicted VMAF scores.

\subsection{Effectiveness of Compression Statistics}
We also studied the performance of regressors trained with and without compression statistics. From Tables \ref{table:libx265-PLCC_1} and \ref{table:libx265-PLCC_2}, it may be observed that, for each regressor trained on video features, the PLCC values between the true VMAF scores and the predicted VMAF scores were greatly improved by employing the compression statistics during training. The regressors showed excellent correlation scores across all resolutions. It is also worth observing that when no video features were employed, i.e., training a regressor only on metadata and compression statistics, good correlations were still obtained. This is because the compression statistics provide additional features/information that uniquely represent RQ points beyond simply using bitrate and resolution. These results demonstrate the effectiveness of including compression statistics, video features, and metadata when training regressors to predict VMAF scores of compressed videos. We employ these regressors to predict the quality of compressed videos, instead of relying only on a VMAF module.

\subsection{Performance of Per-Shot Bitrate Ladders}
\begin{figure}
    \centering
    \begin{subfigure}[b]{\linewidth}
        \includegraphics[width=\linewidth]{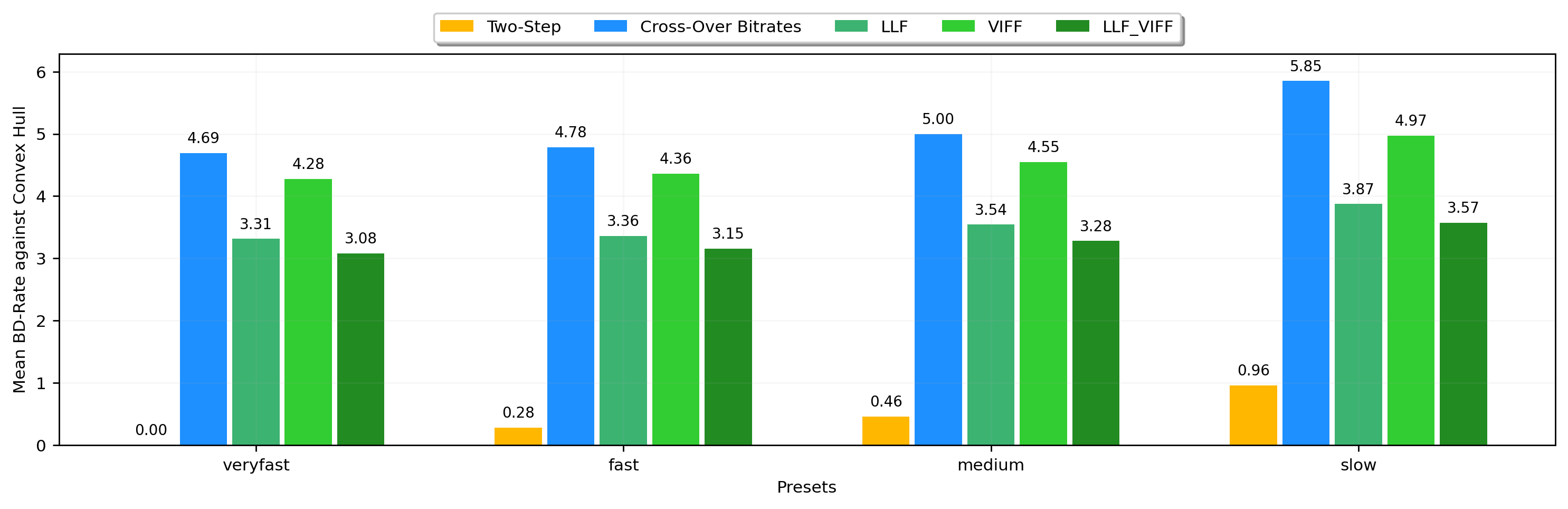}
        \caption{Mean BD-Rate.}
        \label{fig:libx265-libx265_Convex_Hull_1}
    \end{subfigure}
    \hfill
    \begin{subfigure}[b]{\linewidth}
        \includegraphics[width=\linewidth]{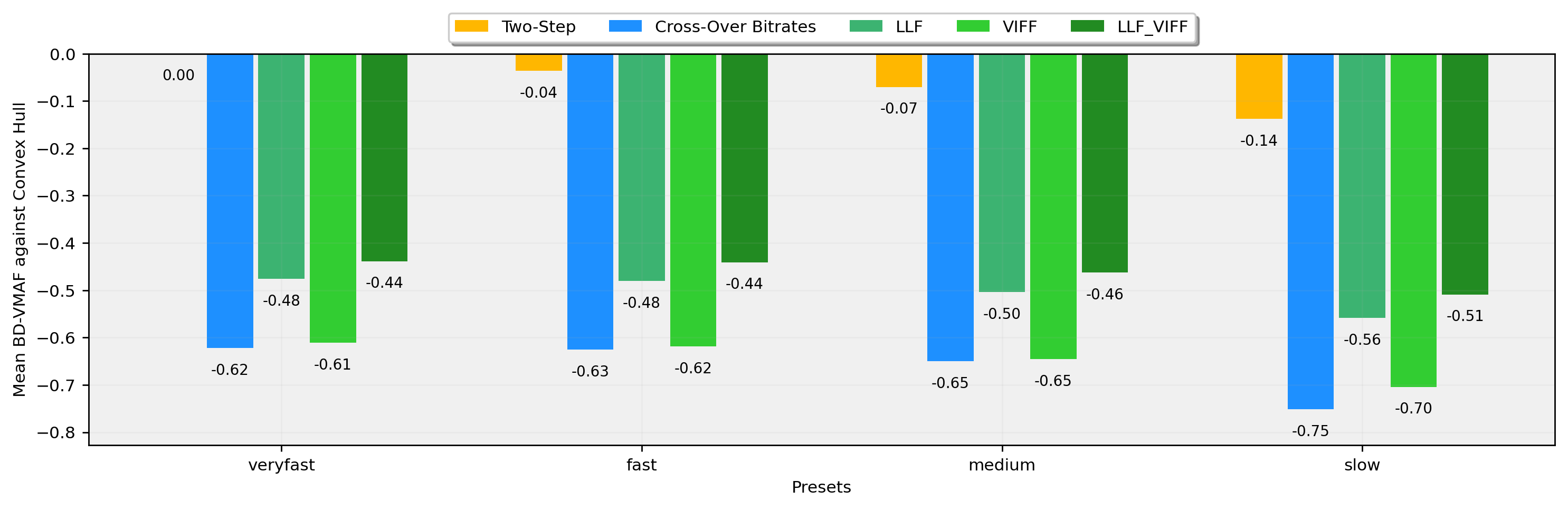}
        \caption{Mean BD-VMAF.}
        \label{fig:libx265-libx265_Convex_Hull_2}
    \end{subfigure}
    \caption{Performances of per-shot bitrate ladders constructed on fast encoder settings against the convex hull, across various presets of the libx265 codec.}
    \label{fig:libx265-libx265_Reference}
\end{figure}

We employed the regressors trained on RQ points from the decided fast encoder settings to predict the per-shot bitrate ladder of each video in our dataset. We evaluated the performance of these predicted per-shot bitrate ladders against the fixed bitrate ladder \cite{Fixed-Bitrate-Ladder} and the convex hull constructed using exhaustive encoding. As described in Section \ref{sec:dataset-experimental-settings}, we tested the performance of these bitrate ladders constructed on the decided fast encoder setting across various encoder settings of the codecs libx265, libsvtav1, libvpx-vp9, and libaom-av1. We plotted performance using the metrics mean BD-rate and mean BD-VMAF against the convex hull, mean BD-rate against the fixed bitrate ladder, and closeness in the form of bar graphs, for each codec. In the figures \ref{fig:libx265-libx265_Convex_Hull_1}, \ref{fig:libx265-libx265_Convex_Hull_2}, gold bars plot the performance of the Two-Step convex hull \cite{Fast-Encoding-Parameter-Selection-for-Convex-Hull-Video-Encoding}, blue bars plot the performance of the cross-over bitrates method \cite{Benchmarking-Learning-based-Bitrate-Ladder-Prediction-Methods-for-Adaptive-Video-Streaming}, and various shades of green bars show the performance of per-shot bitrate ladders constructed using our new model. For better viewing, some of the plots have been moved to the Appendix (Section \ref{sec:appendix}).

\begin{figure}
    \centering
    \begin{subfigure}[b]{\columnwidth}
        \includegraphics[width=\linewidth]{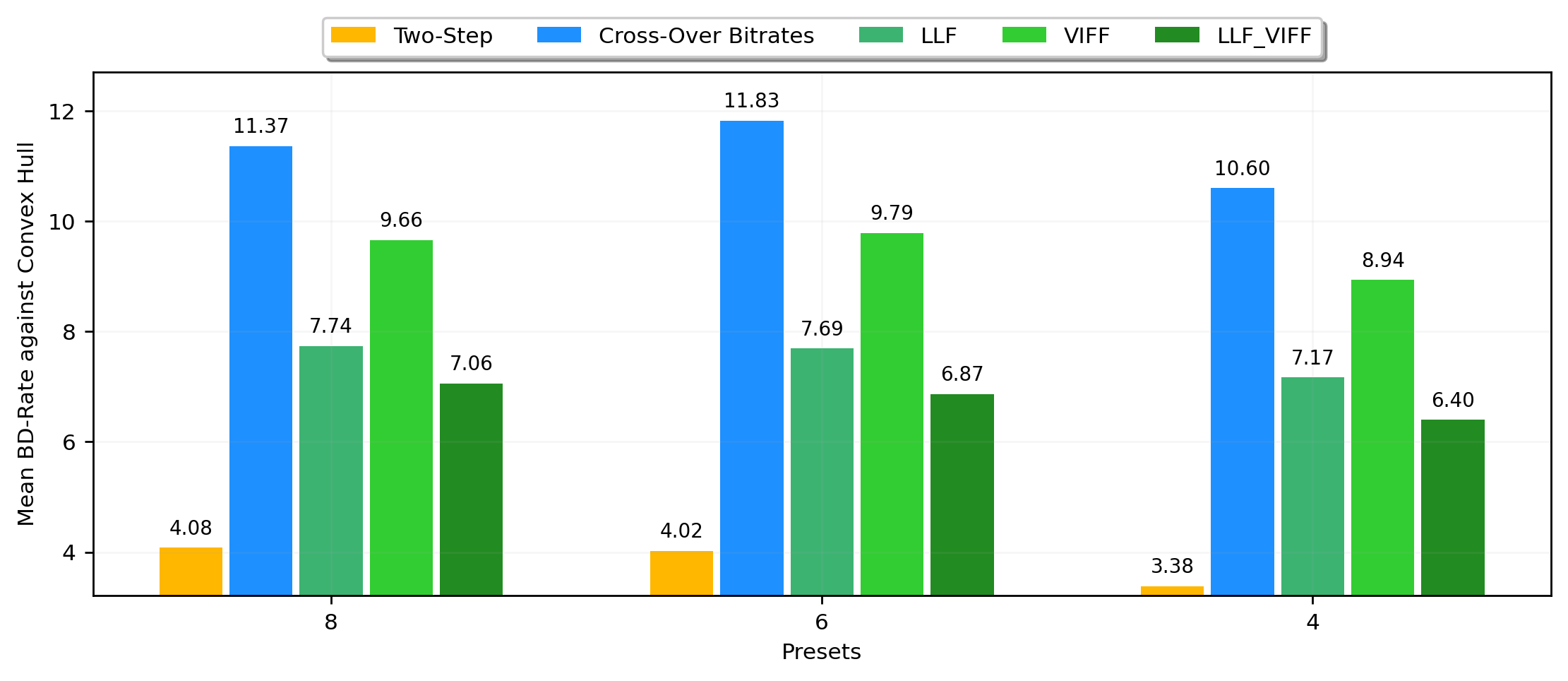}
        \caption{Mean BD-Rate.}
        \label{fig:libx265-libsvtav1_Convex_Hull_1}
    \end{subfigure}
    \hfill
    \begin{subfigure}[b]{\columnwidth}
        \includegraphics[width=\linewidth]{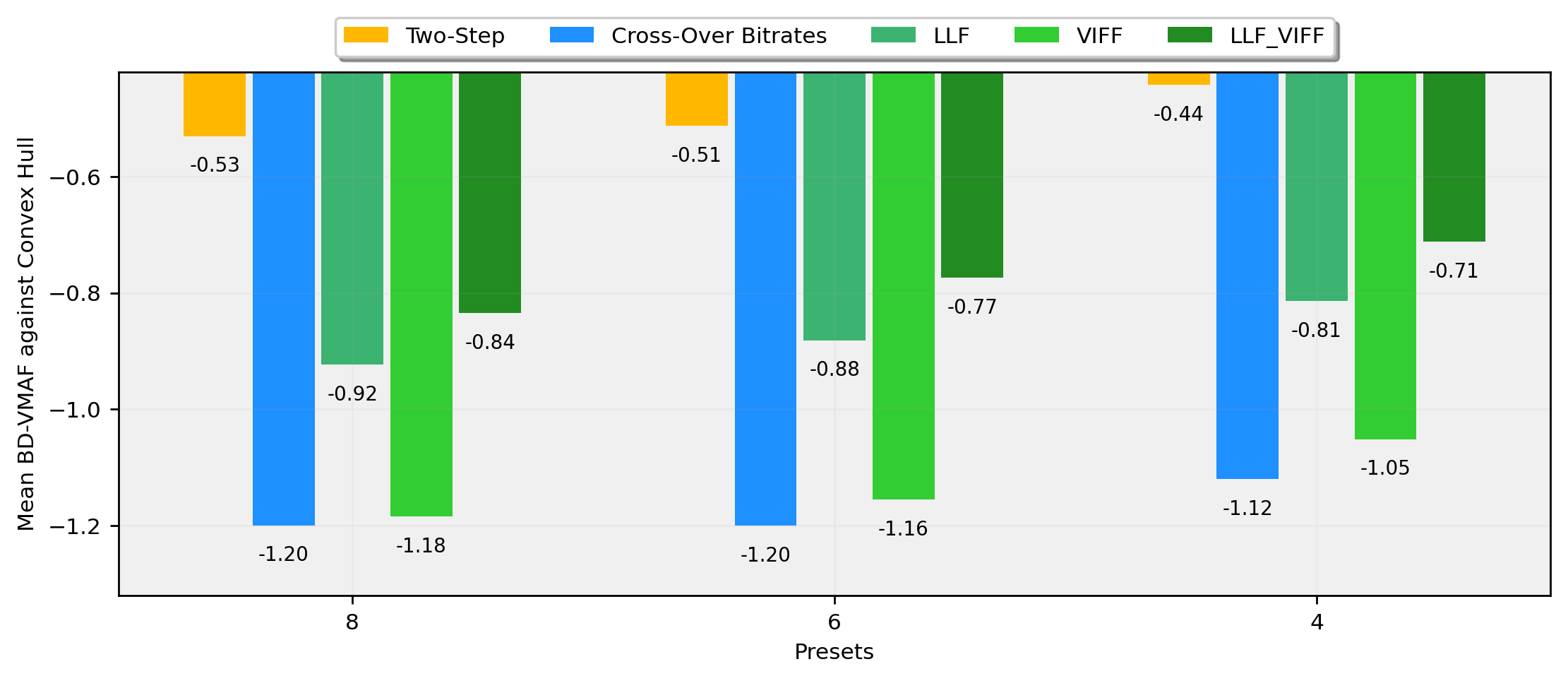}
        \caption{Mean BD-VMAF.}
        \label{fig:libx265-libsvtav1_Convex_Hull_2}
    \end{subfigure}
    \caption{Performances of per-shot bitrate ladders constructed on fast encoder settings against the convex hull, across various presets of the libsvtav1 codec.}
    \label{fig:libx265-libsvtav1_Reference}
\end{figure}

Fig. \ref{fig:libx265-libx265_Reference} shows the performance of per-shot bitrate ladders against the convex hull on various presets of the libx265 codec. Since the bitrate ladders were constructed on the decided fast encoder settings, i.e., libx265 codec with the veryfast preset, the BD-metric loss of the Two-Step convex hull using our fast encoder settings is zero. Fig. \ref{fig:libx265-libx265_Convex_Hull_1} and Fig. \ref{fig:libx265-libx265_Convex_Hull_2} show the mean BD-rate and mean BD-VMAF losses against the convex hull. It may be observed that the Two-Step convex hull method yielded the smallest BD-metric losses, and the loss values gradually increased as the preset became slower. It may be observed that our proposed methods delivered significant decreases in loss as compared to the cross-over bitrates method \cite{Benchmarking-Learning-based-Bitrate-Ladder-Prediction-Methods-for-Adaptive-Video-Streaming}, further demonstrating the effectiveness of our proposed framework and the importance of compression statistics. The reduction in losses is visible across regressors trained on the various video features. Among the regressors trained on the various video features, the regressor trained on low-level features and VIF features obtained the smallest BD-metric losses, followed by low-level features and VIF features, respectively. Across various presets of libx265, our best method delivered a mean BD-rate loss of around 3.3\% and a mean BD-VMAF loss of 0.46 against the convex hull. Figs. \ref{fig:libx265-libx265_Fixed_1} and \ref{fig:libx265-libx265_Fixed_2} in the Appendix show the performance of per-shot bitrate ladders against the fixed bitrate ladder and closeness of the convex hull, respectively. Our proposed model, trained on low-level features and VIF features, also yielded the best gains among the ML-based methods against the fixed bitrate ladder, across the considered set of presets. Based on the closeness values of our proposed method, it may be observed that on around 156 (72\%) of the 217 videos in our video dataset, our best method obtained gains more than 75\% of the gains demonstrated by the convex hull against the fixed bitrate ladder.

\begin{figure}
    \centering
    \begin{subfigure}[b]{0.58\columnwidth}
        \includegraphics[width=\linewidth]{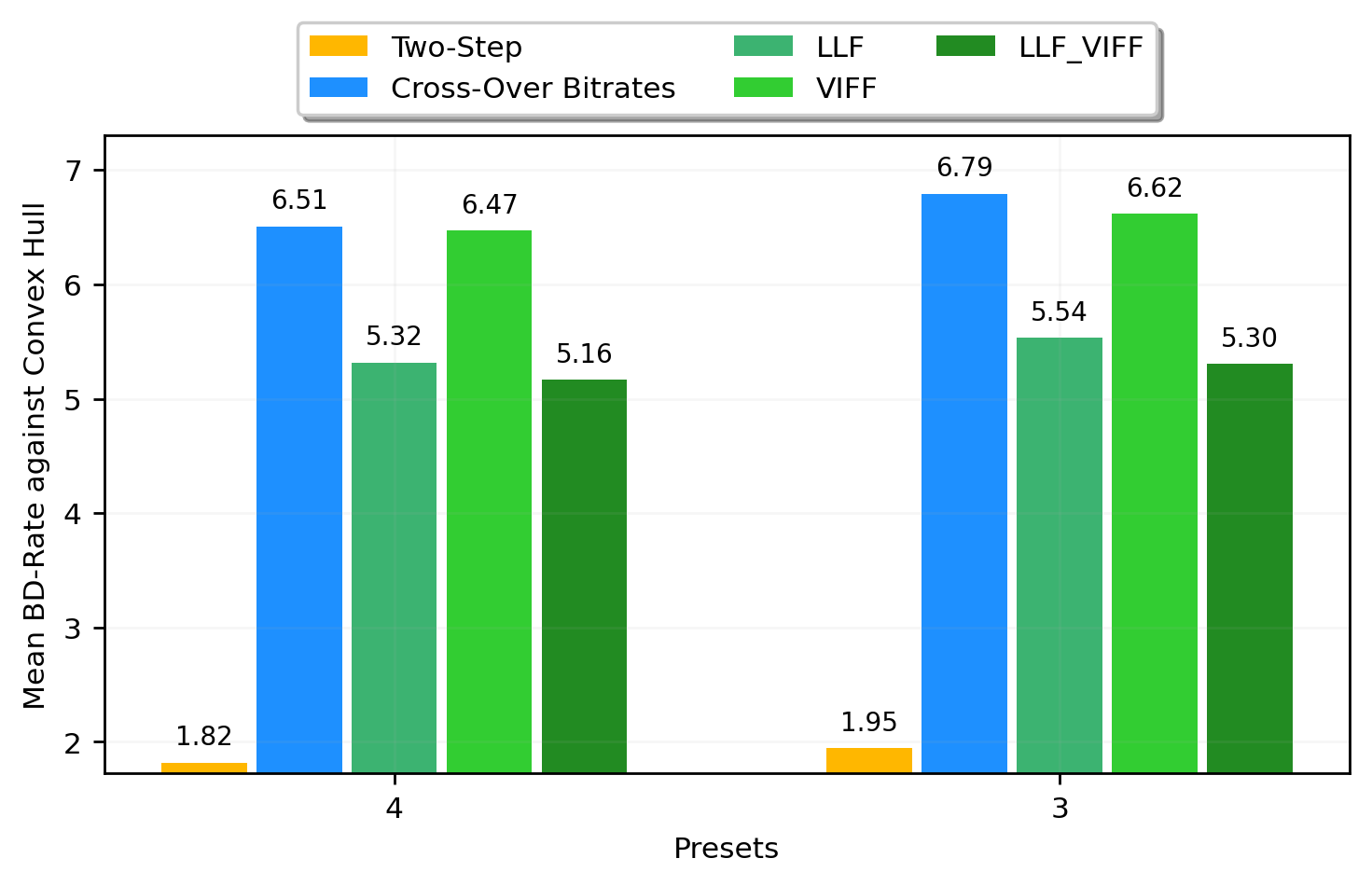}
        \caption{Mean BD-Rate.}
        \label{fig:libx265-libvpx-vp9_Convex_Hull_1}
    \end{subfigure}
    \hfill
    \begin{subfigure}[b]{0.40\columnwidth}
        \includegraphics[width=\linewidth]{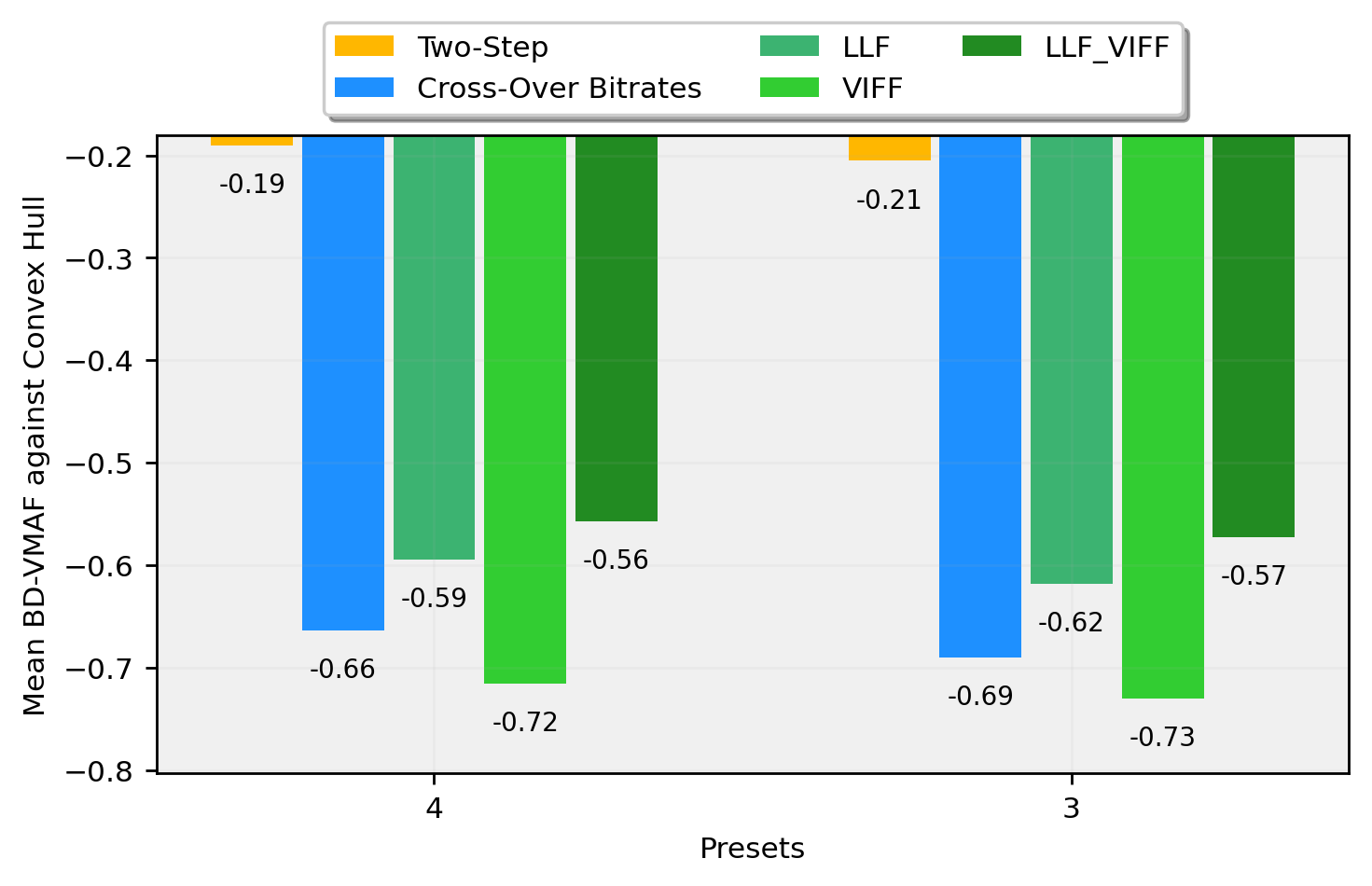}
        \caption{Mean BD-VMAF.}
        \label{fig:libx265-libvpx-vp9_Convex_Hull_2}
    \end{subfigure}
    \caption{Performances of per-shot bitrate ladders constructed on fast encoder settings against the convex hull, across various presets of the libvpx-vp9 codec.}
    \label{fig:libx265-libvpx-vp9_Reference}
\end{figure}

\begin{figure}
    \centering
    \begin{subfigure}[b]{0.58\columnwidth}
        \includegraphics[width=\linewidth]{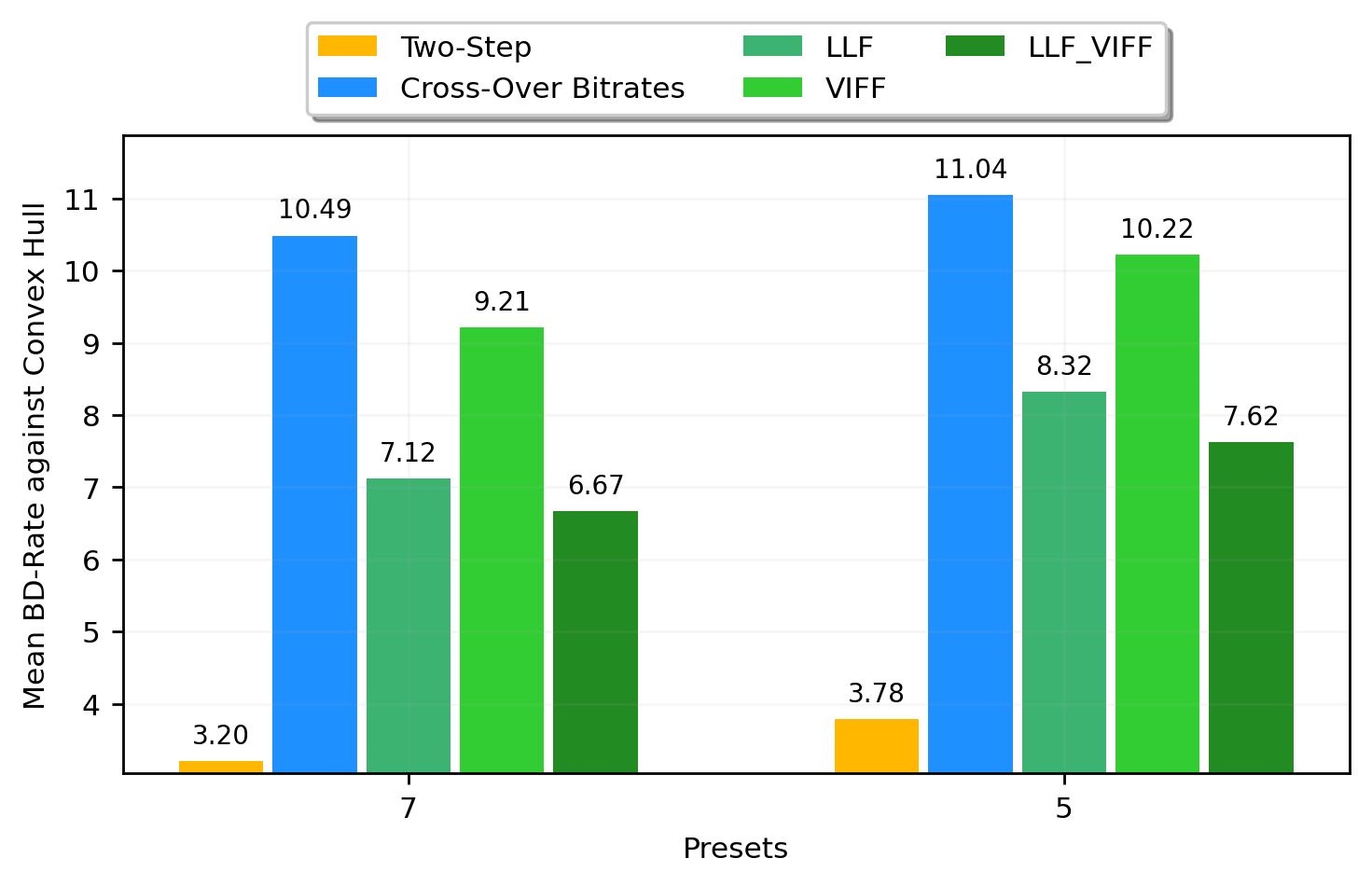}
        \caption{Mean BD-Rate.}
        \label{fig:libx265-libaom-av1_Convex_Hull_1}
    \end{subfigure}
    \hfill
    \begin{subfigure}[b]{0.40\columnwidth}
        \includegraphics[width=\linewidth]{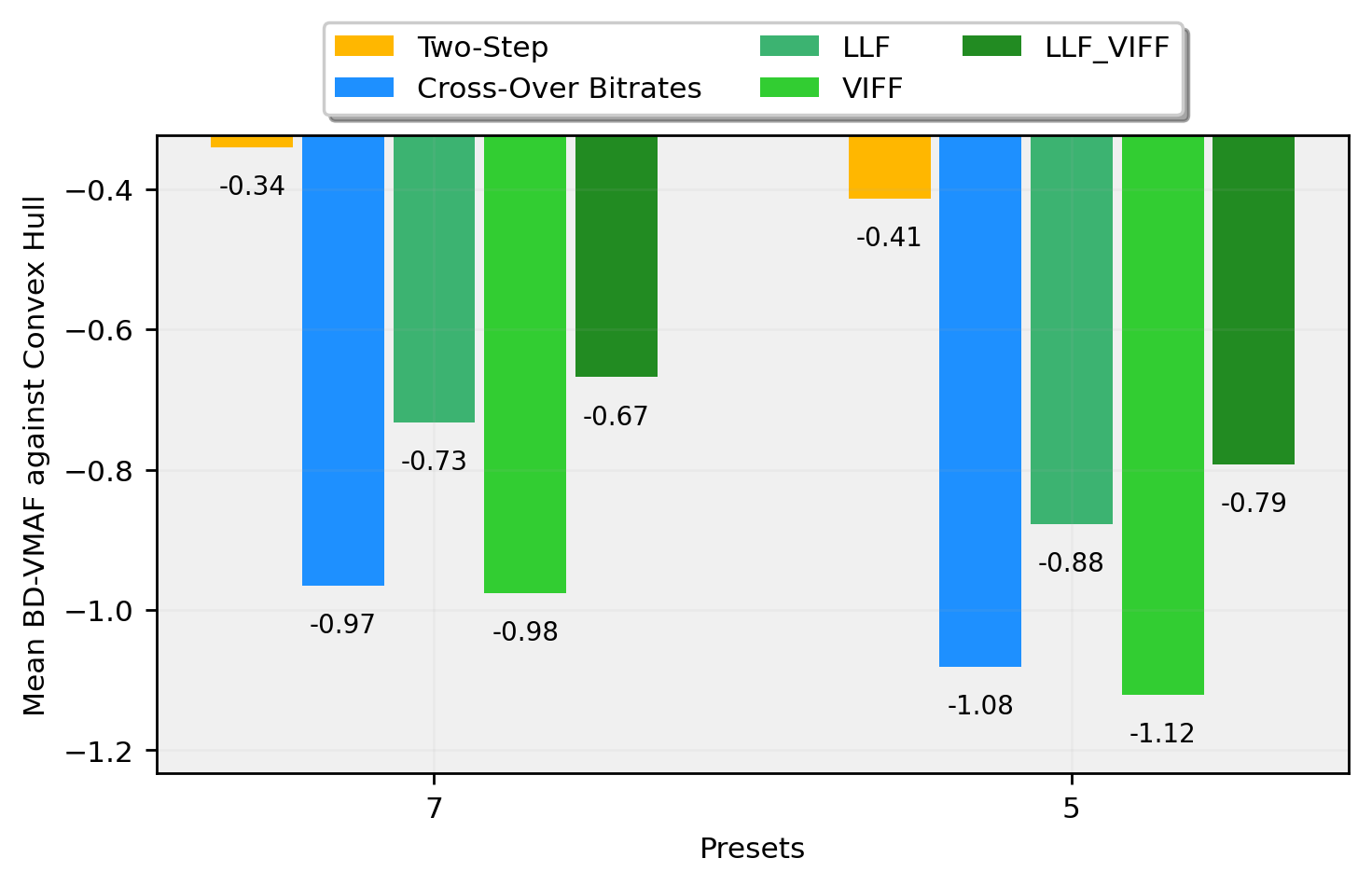}
        \caption{Mean BD-VMAF.}
        \label{fig:libx265-libaom-av1_Convex_Hull_2}
    \end{subfigure}
    \caption{Performances of per-shot bitrate ladders constructed on fast encoder settings against the convex hull, across various presets of the libaom-av1 codec.}
    \label{fig:libx265-libaom-av1_Reference}
\end{figure}

Figures \ref{fig:libx265-libsvtav1_Reference}, \ref{fig:libx265-libvpx-vp9_Reference}, and \ref{fig:libx265-libaom-av1_Reference} show the performance of per-shot bitrate ladders constructed using our fast encoder settings against the convex hull on various encoder settings from the libsvtav1, libvpx-vp9, and libaom-av1 codecs, respectively. Similar to the libx265 codec, the BD-metric losses of the Two-Step convex hull were the smallest among all the compared methods across all encoder settings. Although all three codecs used similar quantization schemes, the constructed per-shot bitrate ladders exhibited smaller losses on libvpx-vp9, as compared to libsvtav1 and libaom-av1. This could be because libx265 and libvpx-vp9 are older codecs, whereas libaom-av1 and libsvtav1 are highly competitive new codecs. Similar to the conclusions obtained on libx265, it may be observed that our new model delivered smaller mean BD-rate and mean BD-VMAF losses than did the cross-over bitrates method \cite{Benchmarking-Learning-based-Bitrate-Ladder-Prediction-Methods-for-Adaptive-Video-Streaming}. Our proposed model trained on the ensemble of low-level features and VIF features gave the best performance among the compared ML-based methods. It may be inferred from these results that the ensemble of our features is more effective than using them separately. On our fast encoder settings using the libx265 codec with the veryfast preset, it may be observed that the performance of the per-shot bitrate ladder constructed on these settings produced larger losses against the respective convex hulls across the libsvtav1, libvpx-vp9, and libaom-av1 codecs. Our best performing method resulted in mean BD-rate losses of 6.7\%, 5.2\%, and 7.1\%, and mean BD-VMAF losses of 0.77, 0.56, and 0.73, on the libsvtav1, libvpx-vp9, and libaom-av1 codecs, respectively.

\begin{table}
    \centering
    \tabcolsep=0.2cm
    \renewcommand{\arraystretch}{1.75}
    \resizebox{\columnwidth}{!}{
    \begin{tabular}{c c c c c c }
    Video-Features & (3840,2160) & (2560,1440) & (1920,1080) & (1280,720) & (960,540) \\
    \hline
    No features & 0.736 & 0.787 & 0.791 & 0.759 & 0.704 \\
    \hline
    $\text{LLF}_{2}$ & 0.87 & 0.878 & 0.878 & 0.859 & 0.847 \\
    \hline
    $\text{VIFF}$ & 0.845 & 0.861 & 0.868 & 0.864 & 0.866 \\
    \hline
    $\text{LLF}_{2}, \text{VIFF}$ & 0.878 & 0.889 & 0.894 & 0.89 & 0.886 \\
    \hline
    \end{tabular}}
    \caption{The average PLCC between true VMAF scores and VMAF scores predicted by Extra-Trees regressors trained on various video features, metadata, and compression statistics from \{libsvtav1, 8\} encoder settings, across all dataset splits.}
    \label{table:libsvtav1:PLCC_1}
\end{table}

\begin{table}
    \centering
    \tabcolsep=0.2cm
    \renewcommand{\arraystretch}{1.75}
    \resizebox{\columnwidth}{!}{
    \begin{tabular}{c c c c c c }
    Video-Features & (3840,2160) & (2560,1440) & (1920,1080) & (1280,720) & (960,540) \\
    \hline
    No features & 0.405 & 0.436 & 0.44 & 0.418 & 0.341 \\
    \hline
    $\text{LLF}_{2}$ & 0.606 & 0.636 & 0.64 & 0.645 & 0.652 \\
    \hline
    $\text{VIFF}$ & 0.514 & 0.549 & 0.564 & 0.622 & 0.658 \\
    \hline
    $\text{LLF}_{2}, \text{VIFF}$ & 0.617 & 0.657 & 0.669 & 0.689 & 0.696 \\
    \hline
    \end{tabular}}
    \caption{The average PLCC between true VMAF scores and VMAF scores predicted by Extra-Trees regressors trained on various video features and metadata from \{libsvtav1, 8\} encoder settings, across all dataset splits.}
    \label{table:libsvtav1:PLCC_2}
\end{table}

Figures \ref{fig:libx265-libsvtav1_Fixed_1}, \ref{fig:libx265-libvpx-vp9_Fixed_1}, and \ref{fig:libx265-libaom-av1_Fixed_1} in the Appendix visualize the performance of per-shot bitrate ladders against the fixed bitrate ladder across the libsvtav1, libvpx-vp9, and libaom-av1 codecs, respectively. Similarly, Figures \ref{fig:libx265-libsvtav1_Fixed_2}, \ref{fig:libx265-libvpx-vp9_Fixed_2}, and \ref{fig:libx265-libaom-av1_Fixed_2} in the Appendix plot the closeness of the predicted per-shot bitrate ladders to the convex hull. It may be observed that across various presets of libsvtav1, libvpx-vp9, and libaom-av1, our proposed methods delivered only 1.5-2.5\% smaller savings than the Two-Step convex hull \cite{Fast-Encoding-Parameter-Selection-for-Convex-Hull-Video-Encoding} against the fixed bitrate ladder. Based on the closeness values, it may be observed that out of 217 videos in our dataset, our best method achieved approximately 75\% savings as demonstrated by the convex hull on 151 (70\%) videos on libsvtav1, 131 (60\%) videos on libvpx-vp9, and 145 (67\%) videos on libaom-av1 codecs. It may also be observed that our models delivered significant improvements to closeness values as compared to the cross-over bitrates method \cite{Benchmarking-Learning-based-Bitrate-Ladder-Prediction-Methods-for-Adaptive-Video-Streaming}. These results demonstrate the effectiveness of our new model and its transferability across various codecs and encoder settings among ML-based methods.

\subsection{Ablation Studies}
The codecs libsvtav1, libvpx-vp9, and libaom-av1 share similar quantization schemes as compared to the considered fast encoder, \textit{viz.} the libx265 codec with veryfast preset. To demonstrate the effectiveness of our model and to understand its transferability as compared to prior work \cite{Benchmarking-Learning-based-Bitrate-Ladder-Prediction-Methods-for-Adaptive-Video-Streaming}, we evaluated the performance of these methods when the libsvtav1 codec with preset 8 was used as a fast encoder. We trained the models on the RQ points of videos compressed using the libsvtav1 codec with preset 8, then used them to predict the per-shot bitrate ladder of each video in the dataset. We evaluated the performance of the predicted per-shot bitrate ladders against the convex hull constructed using exhaustive encoding across various encoder settings of the libx265, libsvtav1, libvpx-vp9, and libaom-av1 codecs. To extract compression statistics during compression when using the SVT-AV1 codec, we slightly altered the existing ffmpeg code and employed ffprobe to extract the relevant statistics. Although AV1 supports B-frames, our analysis with ffprobe showed that videos compressed using the SVT-AV1 codec did not seem to contain any B-frames, which may be related to how the tool reports them. Hence, we fixed the mean and bitrate of B-frames to be -1 in the experiments.

Table \ref{table:libsvtav1:PLCC_1} shows the average PLCC values calculated between the true VMAF scores of compressed videos and the VMAF scores predicted by our proposed quality prediction regressors: Extra-Trees regressors trained on video features, metadata, and compression statistics across all dataset splits. Similar to Table \ref{table:libsvtav1:PLCC_1}, Table \ref{table:libsvtav1:PLCC_2} shows the average PLCC values obtained by our proposed models when compression statistics were not employed: regressors trained only on video features and metadata. The regressors were trained on RQ points computed using libsvtav1 codec preset 8. It may be observed that as compared to using the \{libx265, veryfast\} fast encoder settings, the performance of the regressors trained with and without compression statistics was significantly reduced. These drops in performance arise from two different reasons. First, fewer compression statistics are extracted from the SVT-AV1 codec (four) as compared to those extracted from the libx265 codec (six). Hence, the number of features employed to uniquely represent each RQ point is reduced. Second, it should be noted that as compared to the libx265 codec, the distributions of VMAF scores of videos compressed using the SVT-AV1 codec are heavily skewed towards higher values, as may be observed from Fig. \ref{fig:box_plots_VMAF:libsvtav1_4}. This bias in the dataset could affect the performance of the regressors, as suggested by comparison of the correlations in Tables \ref{table:libx265-PLCC_1} and \ref{table:libsvtav1:PLCC_1}. Although few compression statistics are used, they improve regressor performance, as may be observed by comparing the results in Tables \ref{table:libsvtav1:PLCC_1} and \ref{table:libsvtav1:PLCC_2}.

\begin{figure}
    \centering
    \includegraphics[width=\linewidth]{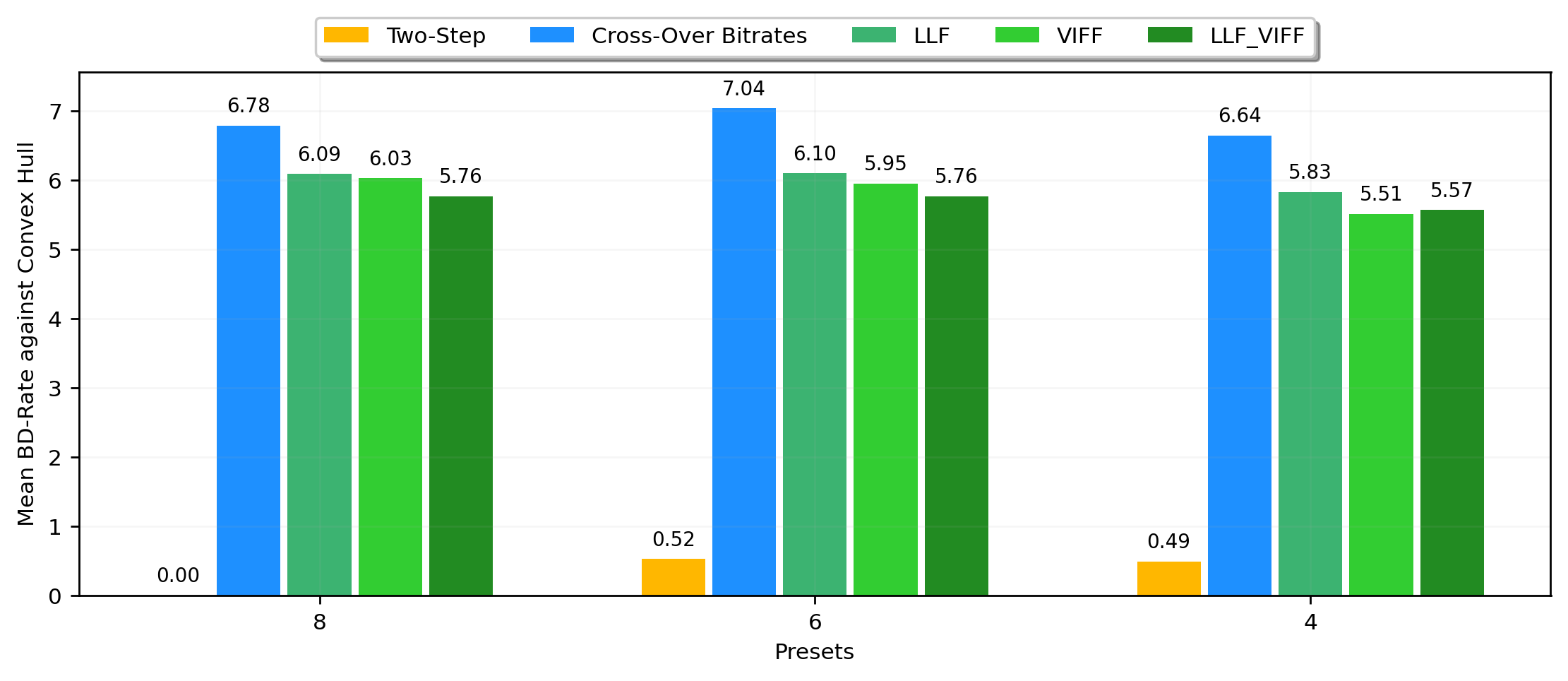}
    \caption{The mean BD-Rate performance of per-shot bitrate ladders constructed on \{libsvtav1, 8\} encoder settings against the convex hull, across various presets of the libsvtav1 codec.}
    \label{fig:libsvtav1-libsvtav1_Reference}
\end{figure}

\begin{figure}
    \centering
    \begin{subfigure}[b]{0.48\columnwidth}
        \includegraphics[width=\linewidth]{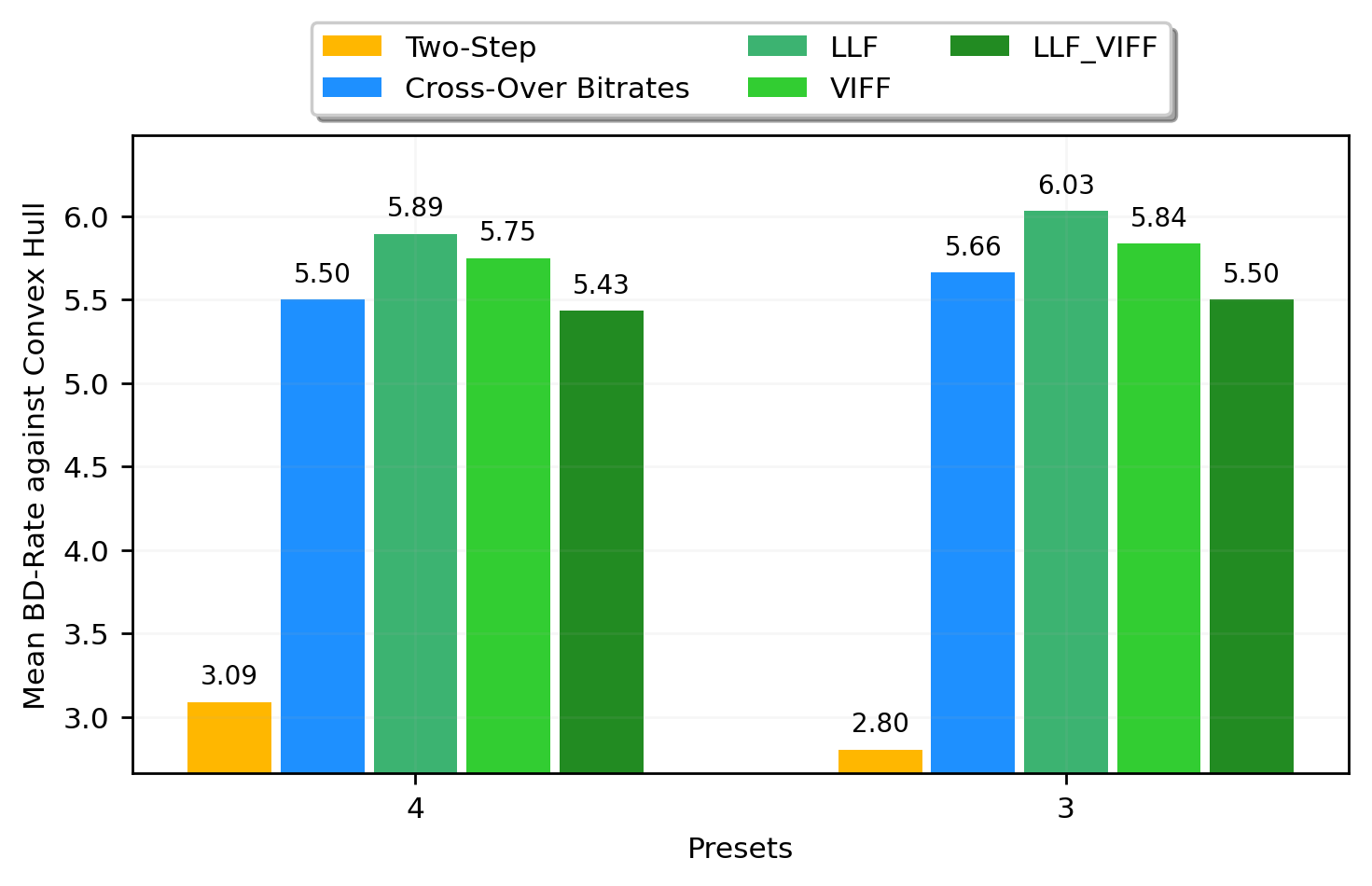}
        \caption{libvpx-vp9.}
        \label{fig:libsvtav1-libvpx-vp9_Reference}
    \end{subfigure}
    \hfill
    \begin{subfigure}[b]{0.48\columnwidth}
        \includegraphics[width=\linewidth]{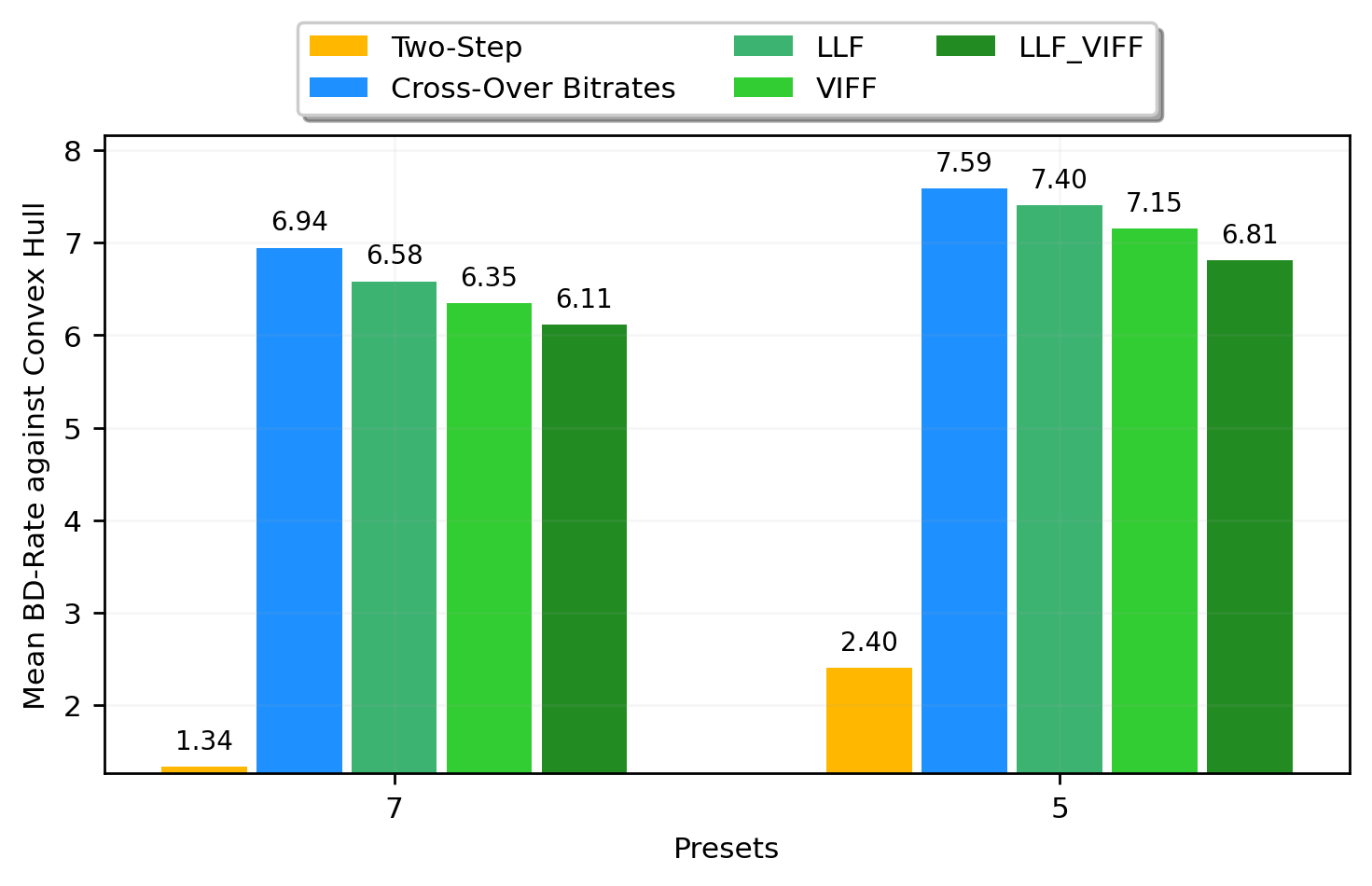}
        \caption{libaom-av1.}
        \label{fig:libsvtav1-libaom-av1_Reference}
    \end{subfigure}
    \caption{The mean BD-Rate performance of per-shot bitrate ladders constructed on \{libsvtav1, 8\} encoder settings against the convex hull, across various presets of the libvpx-vp9 and libaom-av1 codec.}
\end{figure}

\begin{figure}
	\centering
	\includegraphics[width=\linewidth]{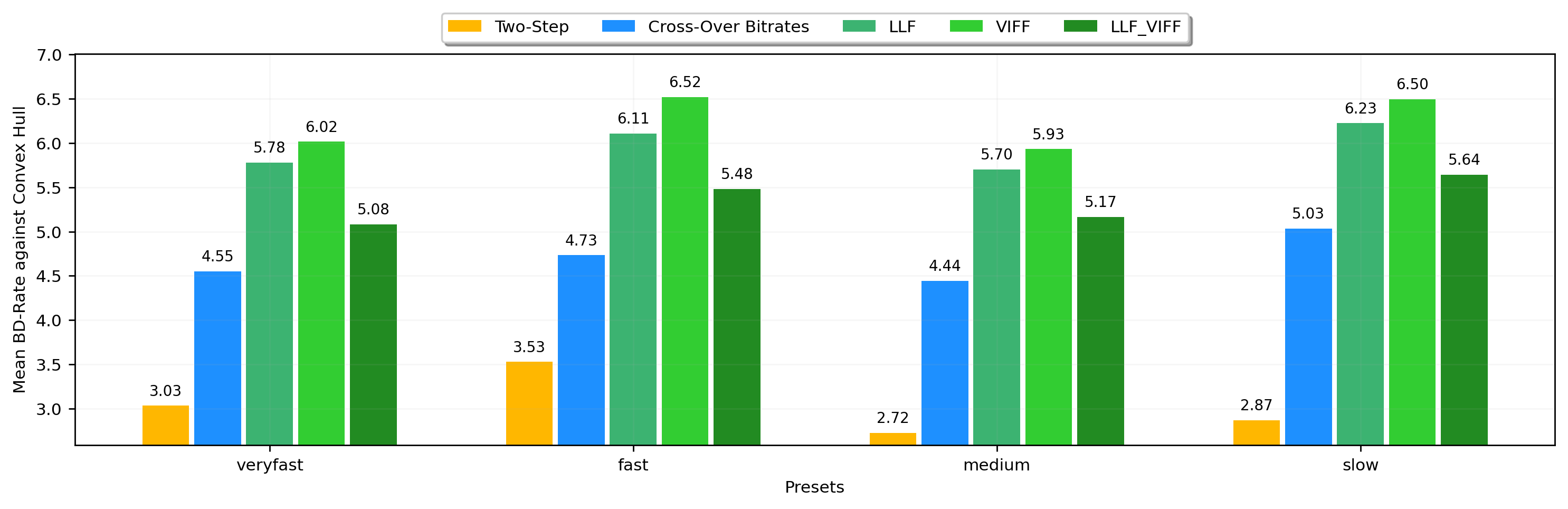}
	\caption{The mean BD-Rate performances of per-shot bitrate ladders constructed on \{libsvtav1, 8\} encoder settings against the convex hull, across various presets of the libx265 codec.}
	\label{fig:libsvtav1-libx265_Reference}
\end{figure}

Figs. \ref{fig:libsvtav1-libsvtav1_Reference}, \ref{fig:libsvtav1-libvpx-vp9_Reference}, and \ref{fig:libsvtav1-libaom-av1_Reference} plot the performance of the predicted per-shot bitrate ladders constructed using the libsvtav1 codec with preset 8 as the fast encoder. As compared to the losses incurred using bitrate ladders constructed using the Two-Step convex hull method \cite{Fast-Encoding-Parameter-Selection-for-Convex-Hull-Video-Encoding} and our model on the fast encoder settings (libx265, veryfast), the losses were substantially lower for the libsvtav1 and libaom-av1 codecs, but are slightly higher for the libvpx-vp9 codec. The loss arising from using the Two-Step convex hull method was substantially higher than from using our proposed model. This disparity in performance can be explained by the legacy nature of the libvpx-vp9 codec. The bitrate ladders constructed using cross-over bitrates resulted in reduced losses as compared to using the initial fast encoder settings across all three codecs. Figure \ref{fig:libsvtav1-libx265_Reference} plots performance across presets of the libx265 codec. As anticipated, when evaluated on the libx265 codec, the performance of the per-shot bitrate ladders constructed using the Two-Step method \cite{Fast-Encoding-Parameter-Selection-for-Convex-Hull-Video-Encoding} and our proposed model using the libsvtav1 codec with preset 8 as the fast encoder settings, were substantially higher than the losses incurred when using the libx265 codec with the veryfast preset as the fast encoder settings. These results show how the performance of predicted bitrate ladders varies against different fast encoder settings and their ability to adapt across various generations of video codecs.

\section{Conclusion}
\label{sec:conclusion}
We developed a novel approach for per-shot bitrate ladder construction by leveraging various video statistics extracted during the compression process. Our new model incorporates features such as the average QPs and bitrates of I, P, and B frames as ``compression statistics'', which we use to train regressors to predict the quality of compressed videos. One advantage of our approach is that it computes the features only once on the source videos, enabling us to achieve substantial computational savings by avoiding the computation of features on upsampled compressed videos when predicting VMAF scores. Our method delivers superior correlations against true VMAF scores, across multiple resolutions and source features in comparison to the prior methods. We employed these trained regressors to construct content-adaptive per-shot bitrate ladders. We evaluated the transferability of the predicted bitrate ladders across different encoder settings spanning four different codecs, by constructing per-shot bitrate ladders on the libx265 codec using the veryfast preset. Our results show that bitrate ladders constructed on fast encoder settings using our new model provide substantially higher bitrate and quality savings as compared to other ML-based methods against the fixed bitrate ladder, across various encoder settings. We also observed reduced losses in bitrate and quality against the convex hull as compared to prior ML-based methods, demonstrating transferability across various encoder settings. Overall, these results demonstrate the efficacy of our new framework for predicting per-shot bitrate ladders, highlighted by substantial bitrate and quality reductions across a wide range of encoder settings, with very minimal computational overhead.

\section*{Acknowledgments}
The authors thank the Texas Advanced Computing Center (TACC) at The University of Texas at Austin for providing HPC resources that have contributed to the research results reported in this paper. URL: http://www.tacc.utexas.edu.

\section*{Appendix}
\label{sec:appendix}

\begin{figure*}
	\begin{subfigure}[b]{0.35\textwidth}
		\includegraphics[width=\linewidth]{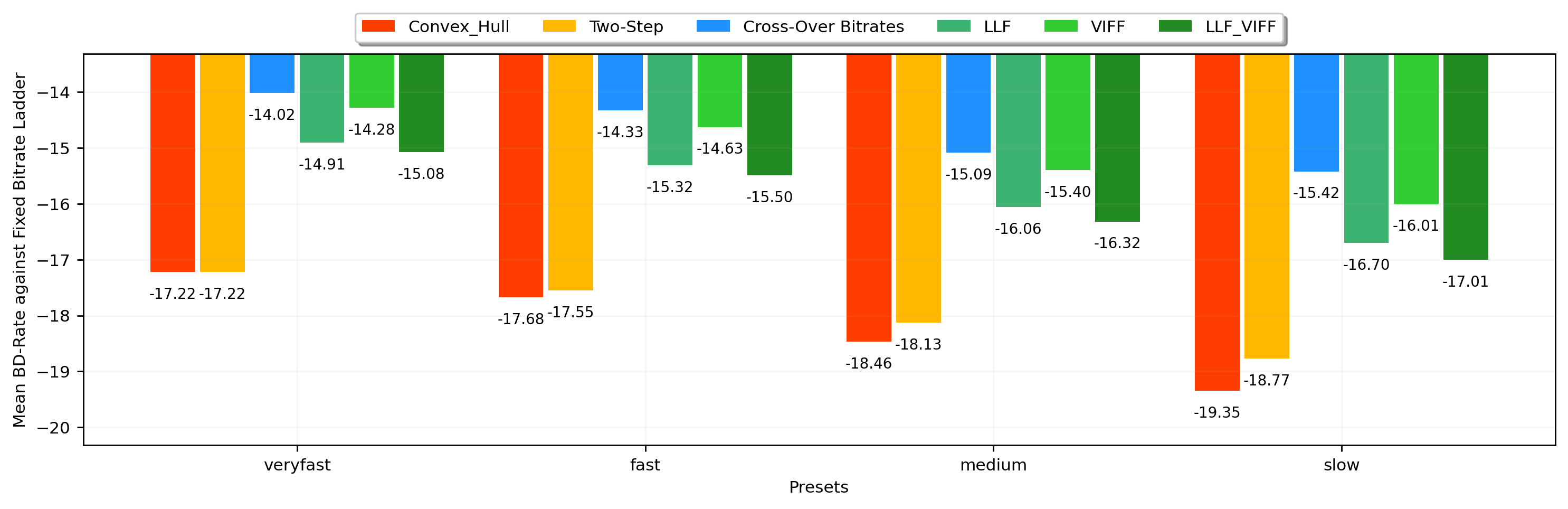}
		\caption{libx265.}
		\label{fig:libx265-libx265_Fixed_1}
	\end{subfigure}
	\hfill
	\begin{subfigure}[b]{0.27\textwidth}
		\includegraphics[width=\linewidth]{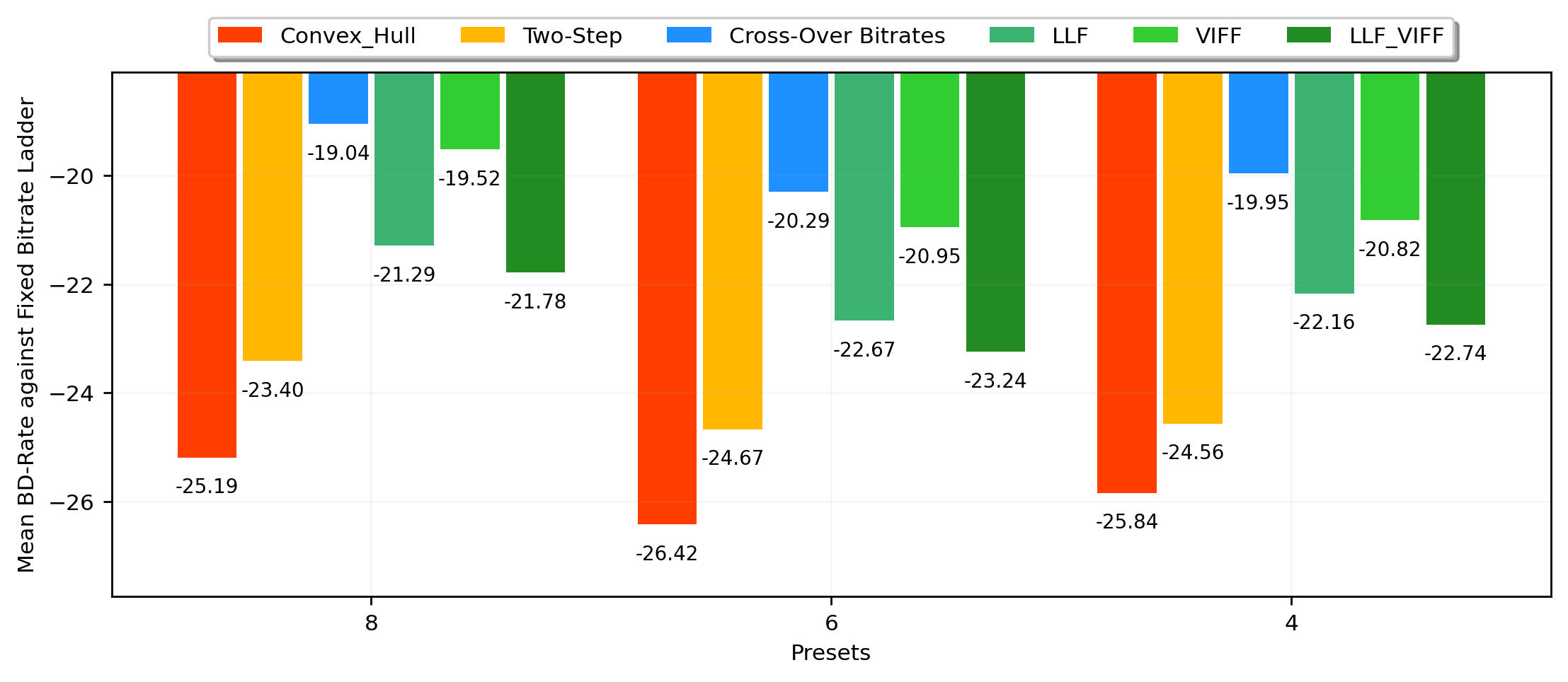}
		\caption{libsvtav1.}
		\label{fig:libx265-libsvtav1_Fixed_1}
	\end{subfigure}
	\hfill
	\begin{subfigure}[b]{0.18\textwidth}
		\includegraphics[width=\linewidth]{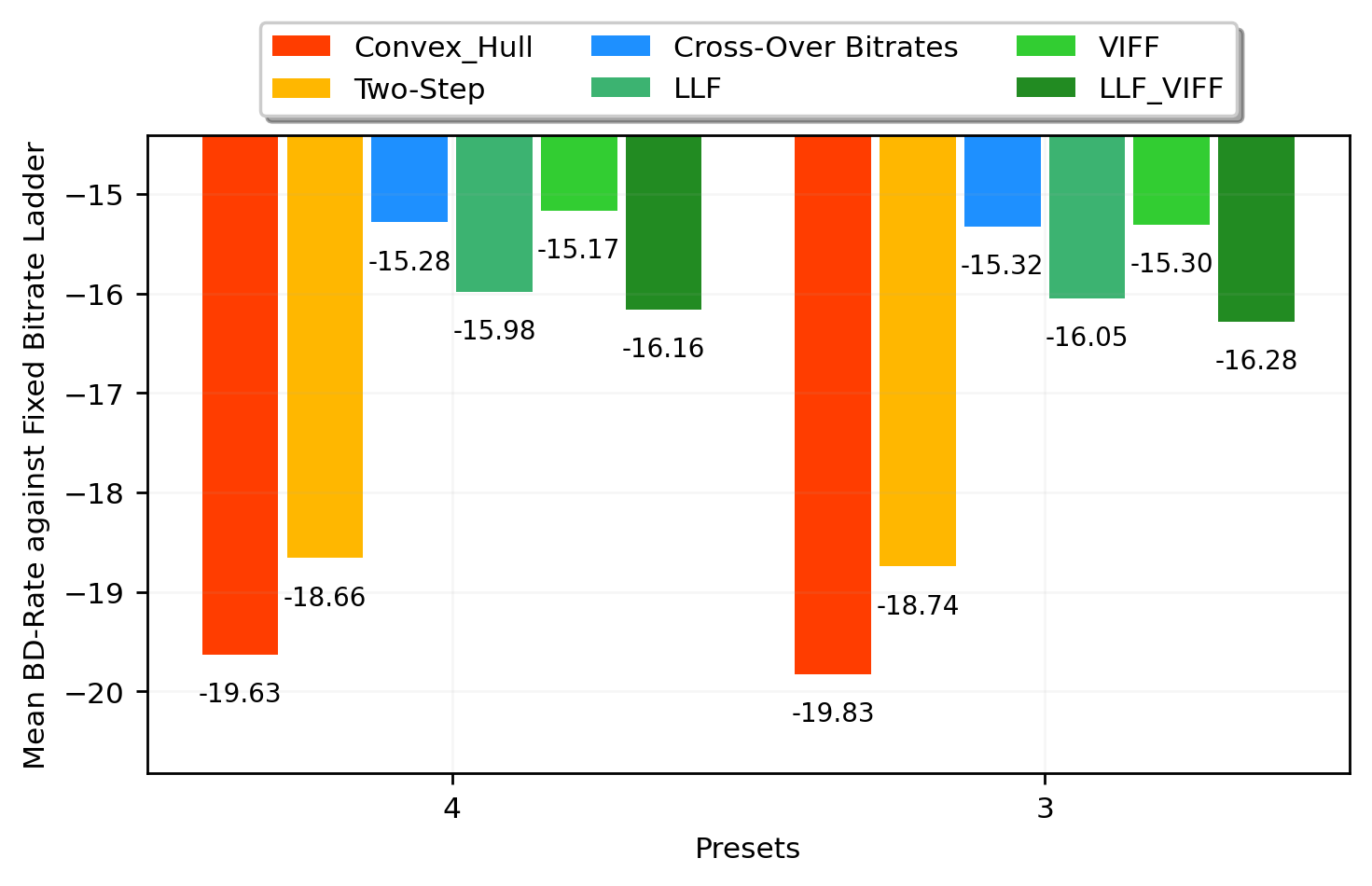}
		\caption{libvpx-vp9.}
		\label{fig:libx265-libvpx-vp9_Fixed_1}
	\end{subfigure}
	\hfill
	\begin{subfigure}[b]{0.18\textwidth}
		\includegraphics[width=\linewidth]{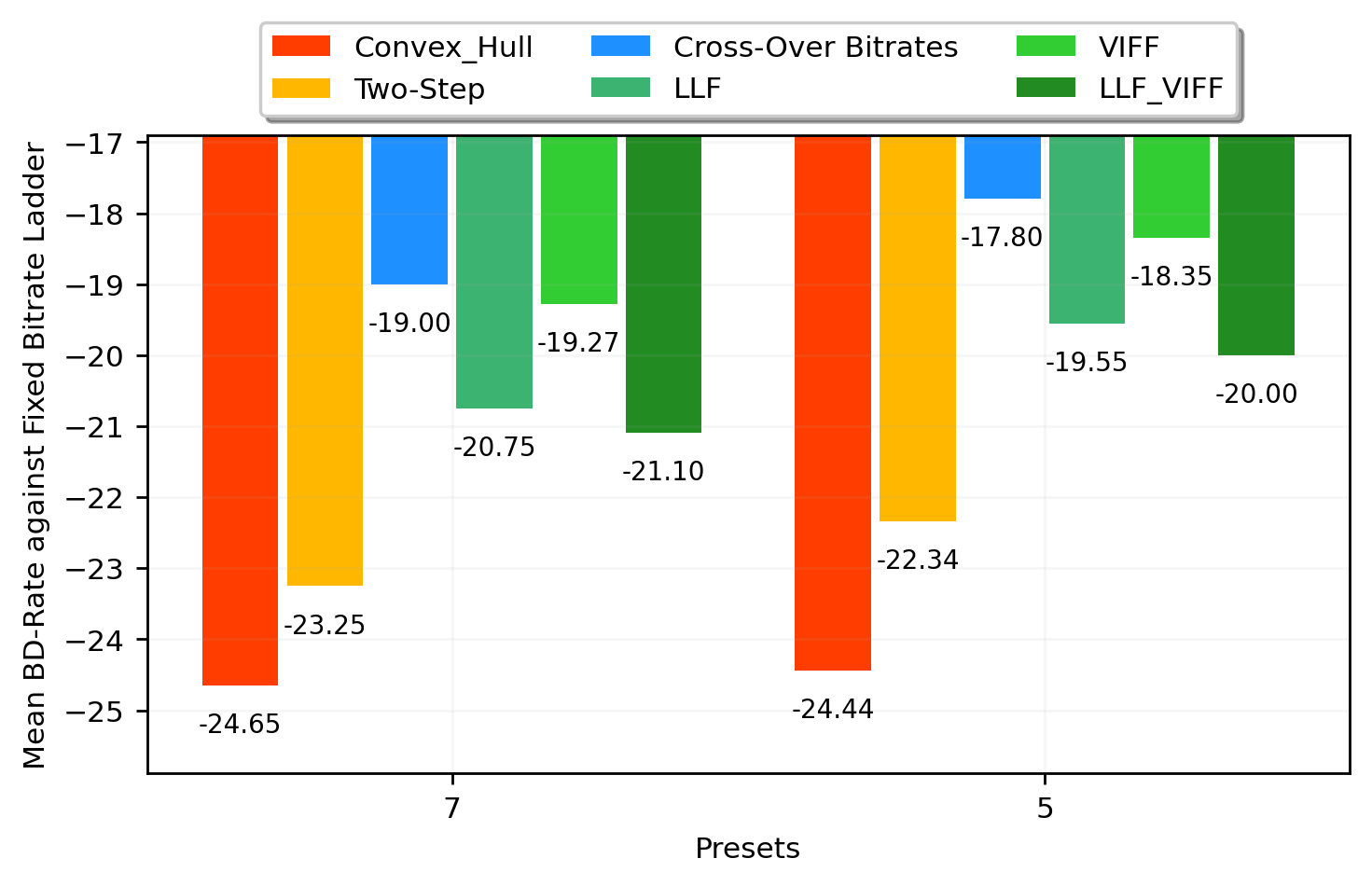}
		\caption{libaom-av1.}
		\label{fig:libx265-libaom-av1_Fixed_1}
	\end{subfigure}
	\caption{Mean BD-Rate performance of per-shot bitrate ladders constructed on fast encoder settings against the fixed bitrate ladder, across various presets of the libx265, libsvtav1, libvpx-vp9, and libaom-av1 codecs.}
	\label{fig:libx265_Fixed}
\end{figure*}

\begin{figure*}
	\begin{subfigure}[b]{0.35\textwidth}
		\includegraphics[width=\linewidth]{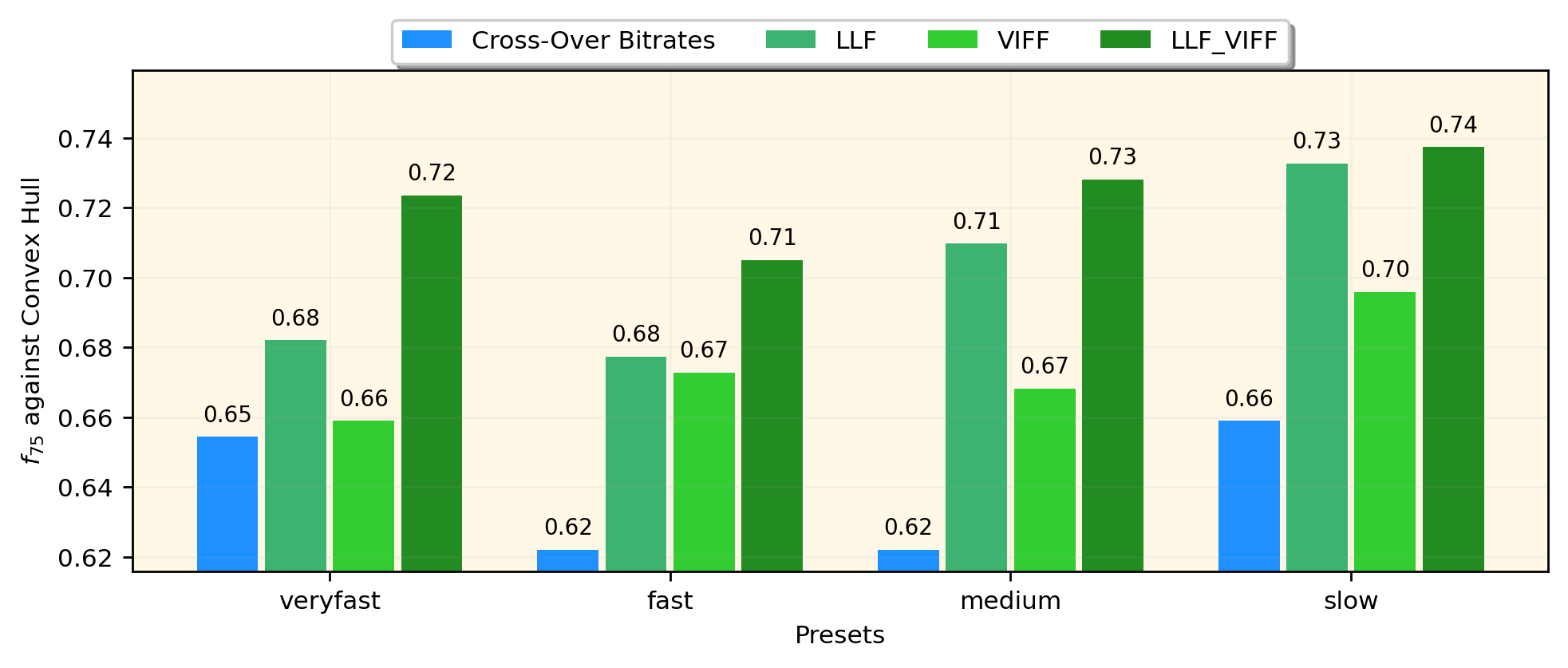}
		\caption{libx265.}
		\label{fig:libx265-libx265_Fixed_2}
	\end{subfigure}
	\hfill
	\begin{subfigure}[b]{0.27\textwidth}
		\includegraphics[width=\linewidth]{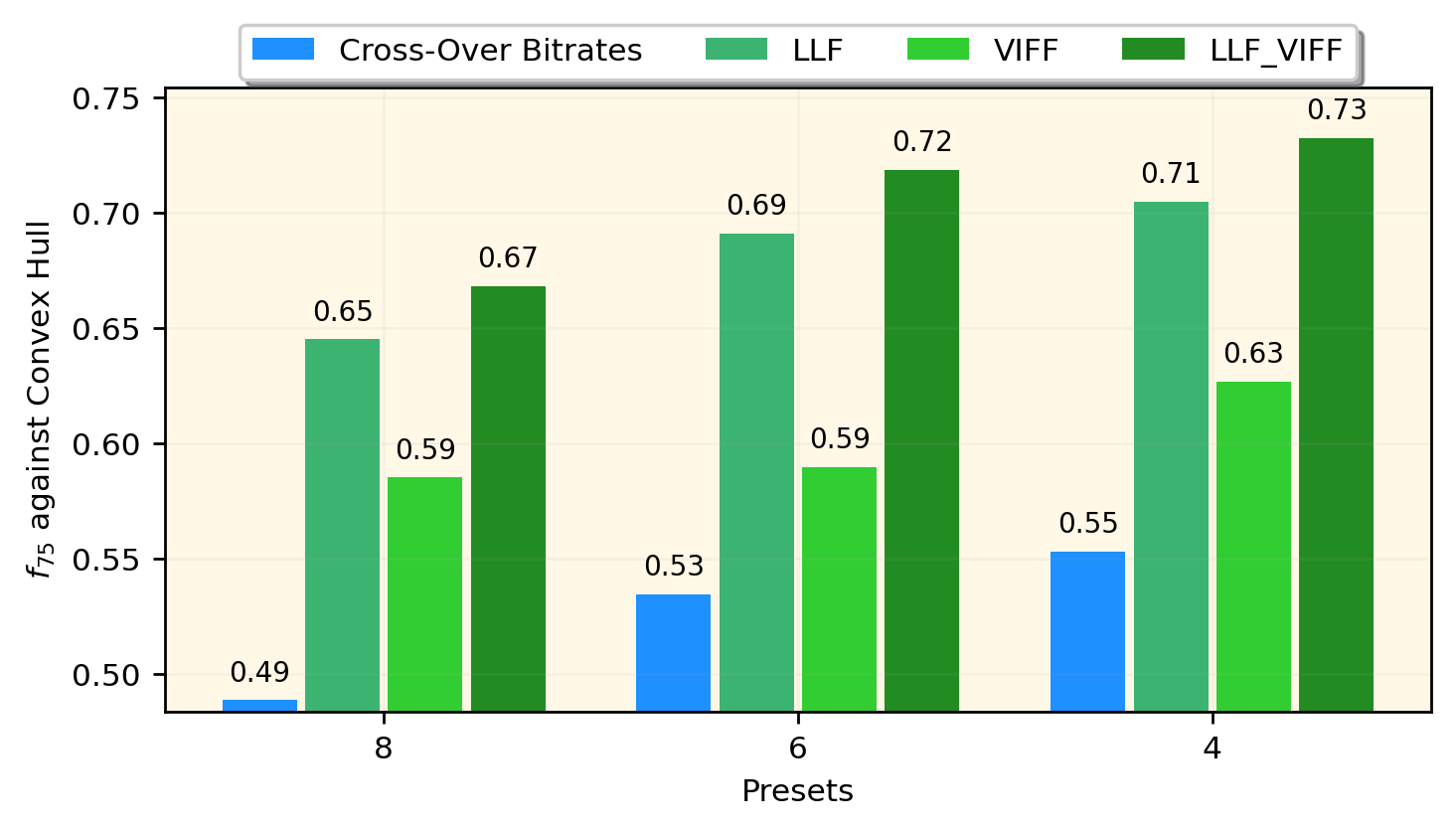}
		\caption{libsvtav1.}
		\label{fig:libx265-libsvtav1_Fixed_2}
	\end{subfigure}
	\hfill
	\begin{subfigure}[b]{0.18\textwidth}
		\includegraphics[width=\linewidth]{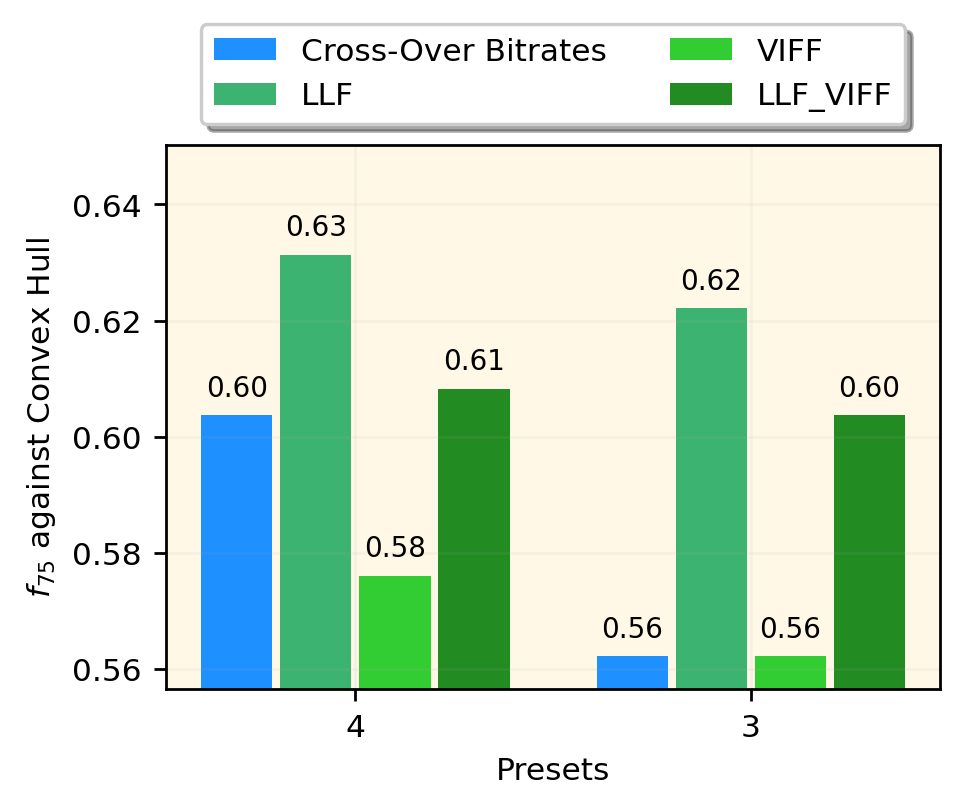}
		\caption{libvpx-vp9.}
		\label{fig:libx265-libvpx-vp9_Fixed_2}
	\end{subfigure}
	\hfill
	\begin{subfigure}[b]{0.18\textwidth}
		\includegraphics[width=\linewidth]{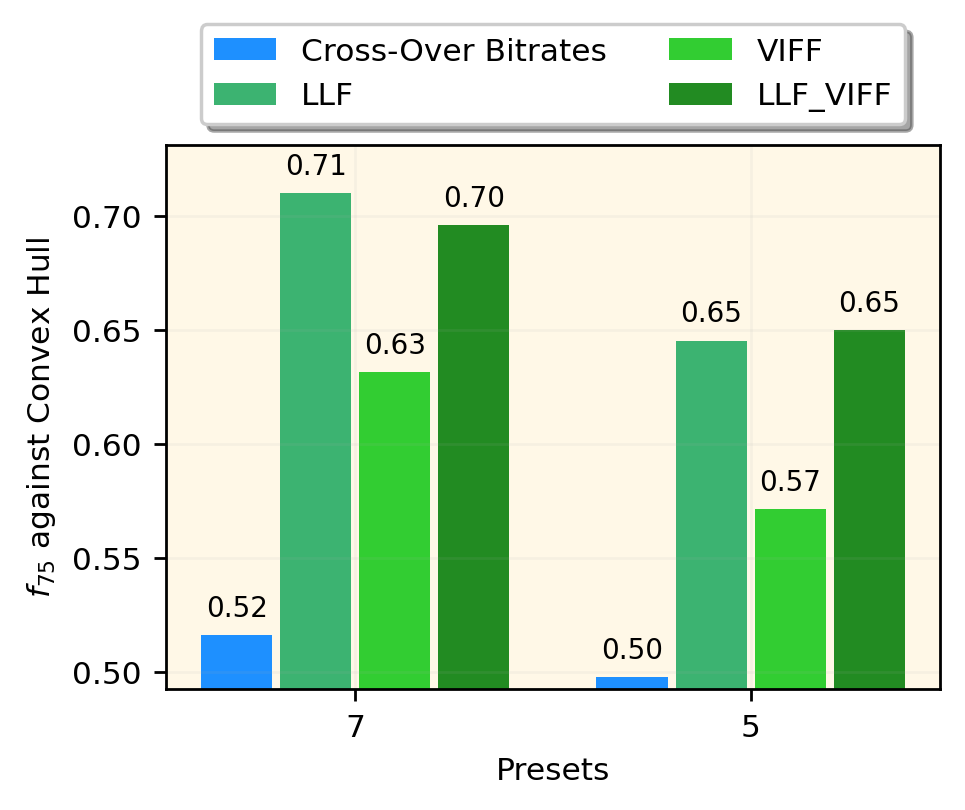}
		\caption{libaom-av1.}
		\label{fig:libx265-libaom-av1_Fixed_2}
	\end{subfigure}
	\caption{Mean $f_{75}$ performance of per-shot bitrate ladders constructed on fast encoder settings against the convex hull, across various presets of the libx265, libsvtav1, libvpx-vp9, and libaom-av1 codecs.}
\end{figure*}

\nocite{*}
\bibliographystyle{IEEEtran}
\bibliography{refs}

\end{document}